\documentclass[twocolumn,trackchanges]{aastex701}

\usepackage{textcomp}
\usepackage{amssymb}
\usepackage{amsmath}
\usepackage{comment}
\usepackage{wasysym}
\usepackage{verbatim}
\usepackage{enumerate}
\usepackage{caption}
\usepackage{morefloats}
\usepackage{multirow}
\usepackage{hyperref}
\usepackage{CJK}
\usepackage{placeins}

\begin{document}
\begin{CJK*}{UTF8}{gbsn}

\title{VLT/MUSE Study of Close AGN Pairs and Host Galaxies in the Local Universe. I. Overview of the Ionized Gas}

\correspondingauthor{Xiaoyu Xu, Zhiyuan Li}
\email{xuxy95@nju.edu.cn, lizy@nju.edu.cn}

\author[0000-0003-0970-535X]{Xiaoyu Xu (许啸宇)}
\affiliation{School of Astronomy and Space Science, Nanjing University, Nanjing 210023, China}
\affiliation{Key Laboratory of Modern Astronomy and Astrophysics, Nanjing University, Nanjing 210023, China}
\email{xuxy95@nju.edu.cn}  

\author[0000-0003-0355-6437]{Zhiyuan Li}
\affiliation{School of Astronomy and Space Science, Nanjing University, Nanjing 210023, China}
\affiliation{Key Laboratory of Modern Astronomy and Astrophysics, Nanjing University, Nanjing 210023, China}
\email{lizy@nju.edu.cn}  

\author[0000-0001-9062-8309]{Meicun Hou}
\affiliation{Institute of Science and Technology for Deep Space Exploration, Suzhou Campus, Nanjing University, Suzhou 215163, China}
\email{houmc@nju.edu.cn}  

\author[0000-0003-4874-0369]{Junfeng Wang}
\affiliation{Department of Astronomy, Xiamen University, Xiamen, Fujian 361005, China}
\email{jfwang@xmu.edu.cn}

\author[0000-0002-1620-0897]{Fuyan Bian}
\affiliation{European Southern Observatory, Alonso de Cordova 3107, Casilla 19001, Vitacura, Santiago 19, Chile}
\email{fbian@eso.org}

\author[0000-0003-3226-031X]{Yan-Mei Chen}
\affiliation{School of Astronomy and Space Science, Nanjing University, Nanjing 210023, China}
\affiliation{Key Laboratory of Modern Astronomy and Astrophysics, Nanjing University, Nanjing 210023, China}
\email{chenym@nju.edu.cn}
\begin{abstract}

Studying AGN pairs and their host galaxies is essential for understanding the interplay between galaxy mergers and key internal processes such as supermassive black hole fueling and feedback.
We cross-match between the Big Multi-AGN Catalog \citep[The Big MAC;][]{2025ApJS..281...25P} and the public data archive of the VLT/MUSE, and obtain 12 AGN pair candidates in the local universe ($z\lesssim0.1$) with a projected distance $r_{\rm p}\leq 20\rm\,kpc$.
Using the archival VLT/MUSE data, we present a spatially resolved study of the ionized gas kinematics and ionization properties of these 12 AGN pair candidates.
By decomposing the optical emission lines into two Gaussian components, we try to separate gas associated with disk rotation from non-circular motions.
We further identify dominant ionization mechanisms using spatially resolved BPT diagnostics. 
We find that both nuclei in 4 of the 12 systems are classified as Seyfert or LINER. 
In addition, three nuclei are classified as star-forming or composite in the optical diagnostics, but are identified as AGNs at other wavelengths. 
Kinematically, regularly rotating ionized gas disks are detected in 16 of 24 nuclei. 
Prominent tidal features traced by ionized gas are also detected in 9 systems.
Ionized gas outflows are widespread and are detected in 18 nuclei. 
Finally, for three nuclei (Mrk 739A, NGC 7592B, and J1544+0446A), we find evidence for fading AGN activity over the past several $10^{4}\rm\, yr$, based on optical emission-line ratios and an assumed AGN photoionization model.

\end{abstract}

\keywords{Galaxies(573) --- Interacting galaxies(802) --- Galaxy winds (626) --- Interstellar medium (847)}


\section{Introduction} \label{sec:intro}

Galaxy mergers are fundamental drivers of galaxy evolution and mass assembly across cosmic time \citep[e.g.][]{2005Natur.435..629S,2006ApJS..163....1H}. 
These events drive gas into galactic centers, triggering starbursts and fueling supermassive black holes (SMBHs) to power active galactic nuclei (AGNs) \citep[e.g.][]{1988ApJ...325...74S,1991ApJ...370L..65B,1996ApJ...471..115B,2005Natur.433..604D,2006ApJS..163....1H}. 
Consequently, compared to isolated galaxies, merging systems are frequently observed to host centrally enhanced star formation (SF), enhanced AGN activity, or dual AGNs \citep[e.g.][]{2011MNRAS.418.2043E,2011ApJ...737..101L,2012ApJ...745...94L,2012ApJ...746L..22K,2014MNRAS.441.1297S,2015MNRAS.447.2123C}.

During a merger, two SMBHs eventually spiral inward and form a bound binary, and harden before finally coalescing \citep[e.g.][]{1980Natur.287..307B}. 
This coalescence generates powerful gravitational waves, offering a unique way to test general relativity and probe SMBH populations in the early universe \citep[][]{1994MNRAS.269..199H,2005ApJ...629...15H,2016PhRvD..93b4003K}. However, observing bound binaries at pc scales remains difficult and rare \citep[e.g.][]{2012NewAR..56...74P,2022LRR....25....3B}.

Although gravitationally bound systems are elusive, the earlier evolutionary phase of kpc-scale dual AGNs provides a direct window into the merger-driven co-evolution of black holes and galaxies. 
Consequently, extensive efforts have been dedicated to identifying and characterizing these systems through a variety of complementary approaches.
Systematic searches using galaxy pairs with available fiber spectra for both nuclei have revealed simultaneous AGN activity \citep[e.g.][]{2011MNRAS.418.2043E,2011ApJ...737..101L,2012ApJ...745...94L}. 
High-resolution X-ray imaging with {\it Chandra} has proven particularly powerful in spatially resolving dual AGNs at sub-arcsecond scales \citep[e.g.][]{2003ApJ...582L..15K,2010RAA....10..309W,2012ApJ...746L..22K}. 
Mid-infrared selection techniques using data from the Wide-field Infrared Survey Explorer (WISE) have further uncovered a large population of dual AGN candidates by exploiting the distinct infrared signatures of AGN activity. 
Radio interferometry has likewise contributed significantly to this effort, detecting compact dual cores or morphologically distinct jet structures that are characteristic signatures of actively accreting black holes \citep[e.g.][]{2015ApJ...799...72F,2015ApJ...813..103M,2021AJ....162..289Z,2021AJ....162..276Z}.

To explain observed luminosity functions and the tight scaling relations between SMBH mass and host bulge properties \citep[e.g. the $M_{\rm BH}$--$\sigma_{*}$ relation;][]{2013ARA&A..51..511K,2013ApJ...764..184M}, feedback processes are required to self-regulate galaxy growth \citep[e.g.][]{1998A&A...331L...1S, 2012ARA&A..50..455F,2015ARA&A..53..115K}.
Both SF and AGN activity launch kpc-scale winds \citep[e.g.][]{2005ARA&A..43..769V,2008ApJS..175..356H,2014ApJ...781...55W,2017MNRAS.472L.109A}.
In nearby galaxies, especially mergers, observations show multiphase SF winds that extend several kpc and reach velocities of a few hundred $\rm\,km\,s^{-1}$ \citep[e.g.][]{1990ApJS...74..833H,2000ApJS..129..493H,2005ApJS..160..115R,2013ApJ...768...75R}.
Similarly, multiphase AGN outflows have been clearly identified across a wide range of scales and redshifts \citep[e.g.][]{2012ARA&A..50..455F,2021NatAs...5...13L,2024Galax..12...17H}.
By interacting with the interstellar, circumgalactic, and intergalactic media (ISM, CGM, and IGM), AGN outflows are expected to substantially influence the growth and evolution of the host galaxy \citep[][]{1998A&A...331L...1S,2011ApJ...736...62W,2011ApJ...742...23W,2012ARA&A..50..455F,2015ARA&A..53..115K,2022ApJ...938..127X,2024Galax..12...17H}. 
In the negative feedback scenario, outflows heat or expel cold gas, which limits the fuel available for new stars and reduces the star formation rate (SFR) \citep[e.g.][]{1998A&A...331L...1S,2008ApJS..175..356H,2012ARA&A..50..455F,2012RAA....12..917S}.
Alternatively, some studies indicate a positive feedback mechanism where outflows compress the ambient gas and trigger star formation \citep[e.g.][]{2013ApJ...772..112S,2017MNRAS.468.4956Z}.

Theoretically, stellar feedback and AGN feedback are distinct processes. 
However, separating their observational signatures is difficult, especially in the chaotic environments of merging systems. 
For instance, both compact nuclear starbursts and AGNs can drive high-velocity winds \citep[e.g.][]{2005ARA&A..43..769V,2021ApJ...923..275P,2022ApJ...933..110X,2024MNRAS.535.1684T}. 
Consequently, it is often impossible to distinguish between them using spatially integrated spectroscopy alone. 
To identify the primary driver of the feedback, it is essential to spatially resolve the ionization structure and gas kinematics \citep[e.g.][]{2018NatAs...2..176C}.

Integral Field Spectroscopy (IFS) addresses this need by generating a data cube that contains a spectrum for every spaxel in the field of view (FoV). 
This provides detailed maps of the kinematics for both gas and stars. 
It also allows us to map key physical properties such as gas ionization and chemical abundances \citep[e.g.][]{2012A&A...538A...8S,2015MNRAS.446.1567A,2015ApJ...798....7B,2026ApJS..282...16L}.
By combining the spatial emission-line diagnostic diagrams \citep[BPT diagrams;][]{1981PASP...93....5B,1987ApJS...63..295V} with velocity and velocity dispersion fields, it becomes possible to spatially segregate distinct ionization sources and gas components. 
Feedback-driven outflows and gas inflows could be distinguished from the ionized gas disks and tidal features \citep[e.g.][]{2013ApJ...768...75R,2014MNRAS.441.3306H,2019A&A...622A.146M,2020AJ....159..167L,2022ApJ...933..110X,2022MNRAS.511.2105K,2023SciA....9G8287S,2023ApJ...943L..25Z,2025ApJ...993...35X}.

Among current facilities, the Multi Unit Spectroscopic Explorer (MUSE) on the Very Large Telescope (VLT) offers a unique combination of capabilities ideal for this type of study \citep[][]{2010SPIE.7735E..08B}. 
The high sensitivity of MUSE enables the detection of low surface brightness Extended Emission Line Regions (EELRs) which are essential for tracing large-scale feedback and ionization echoes \citep[e.g.][]{2019A&A...622A.146M,2020AJ....159..167L,2023A&A...679A..88Z,2025ApJ...993...35X}.
Furthermore, the 0.\arcsec2 spatial sampling allows to resolve the circumnuclear regions in the local universe, and the large $1\arcmin \times1\arcmin$ FoV ensures that the bulge of the merging system can be covered.

We introduce a comprehensive VLT/MUSE project designed to investigate the spatially resolved properties of close galaxy mergers hosting AGN pair candidates. 
Our sample consists of 12 systems selected from the MUSE archive, all located at $z\lesssim 0.1$ with projected nucleus separations of $r_{\rm p}\leq 20\rm\,kpc$.
In this first paper of the series, we present the spatially resolved surface brightness maps of major optical emission lines along with the detailed gas velocity and velocity dispersion fields for each system. 
By employing spatially resolved BPT diagnostic diagrams, we characterize the dominant ionization mechanisms on a spaxel-by-spaxel basis. 
Our discussion primarily concentrates on diagnosing the ionization mechanisms of the nuclei and the physical origins of the extended ionized gas structures. 
This analysis establishes the observational framework required to identify ionization cones and large-scale outflows. 
It lays the groundwork for the subsequent papers in this series: Paper II will explore the spatially resolved properties of the stellar populations, and Paper III will present a detailed quantitative analysis of the gas outflows.
In Section \ref{sec:sample and data}, we describe the sample and the methodology. 
The main results are presented in Section \ref{sec:result}. 
The discussion and conclusions are given in Sections \ref{sec:discussion} and \ref{sec:summary}, respectively.

\section{Sample selection and data reduction} \label{sec:sample and data}

\subsection{Sample selection} \label{subsec:sample}

\begin{table*}[]
\centering
\begin{tabular}{ccccccc}
Name  & Nucleus  & z      & RA     & Dec    & log $M_{*}$  & Separation  \\
(1) & (2) & (3) & (4) & (5) & (6) & (7)  \\
\hline
\hline
ESO 509-066E       & A & 0.0332 & 13:34:40.76 & $-$23:26:45.35           & 10.0    & 16.\arcsec00      \\
ESO 509-066W       & B & 0.0343 & 13:34:39.62 & $-$23:26:47.45           & 10.4    & 10.93 kpc   \\
\hline
IC 1623 E           & A & 0.0205  & 1:07:47.62  & $-$17:30:24.28        & 10.4    &15.\arcsec05        \\
IC 1623 W           & B & 0.0205  & 1:07:46.55  & $-$17:30:22.29        & 10.1    &6.24 kpc     \\
\hline
Mrk 463E            & A & 0.0508 & 13:56:02.88 & +18:22:18.44        & 11.6    &3.\arcsec83       \\
Mrk 463W            & B & 0.0504   & 13:56:02.61 & +18:22:17.88        & 10.9    &3.77 kpc  \\
\hline
Mrk 739E            & A & 0.0299  & 11:36:29.36 & +21:35:46.07        & 10.8    &5.\arcsec80       \\
Mrk 739W            & B & 0.0298 & 11:36:28.93 & +21:35:46.59        & 10.3    &3.47 kpc  \\
\hline
NGC 1128N          & A & 0.0222 & 2:57:41.56  & +06:01:36.97        & 10.6     & 16.\arcsec16   \\
NGC 1128S          & B & 0.0238 & 2:57:41.64  & +06:01:20.65        & 10.3     & 7.77 kpc     \\
\hline
NGC 3341A           & A & 0.0273 & 10:42:31.46 & +05:02:37.79         & 10.3  & 9.\arcsec6   \\
NGC 3341B           & B & 0.0274 & 10:42:32.05 & +05:02:42.13         & 9.7   & 5.3 kpc  \\
\hline
NGC 7592E          & A & 0.0244 & 23:18:22.65 & $-$04:24:59.06        & 10.2  & 10.\arcsec96     \\
NGC 7592W          & B & 0.0246 & 23:18:21.79 & $-$04:24:56.64           & 10.2  &5.44 kpc  \\
\hline
SDSS J005114.10+002049.5  & A & 0.113   & 0:51:14.11  & +00:20:49.41       & 11.0   &3.\arcsec40          \\
SDSS J005113.92+002047.0 & B & 0.112 & 0:51:13.94  & +00:20:47.08        & 10.8  &6.96 kpc     \\
\hline
SDSS J085312.70+162615.5 & A & 0.0649   & 8:53:12.85  & +16:26:16.19        & 10.3  & 6.\arcsec32     \\
SDSS J085312.36+162619.5 & B & 0.0637 & 8:53:12.35  & +16:26:19.76        & 10.6  &7.88 kpc  \\
\hline
SDSS J085837.68+182223.4 & A & 0.0589   & 8:58:37.66  & +18:22:23.56        & 10.9  &2.\arcsec79      \\
SDSS J085837.53+182221.6 & B & 0.0587   & 8:58:37.52  & +18:22:21.64         & 10.8  &3.18 kpc  \\
\hline
SDSS J141447.48-000011.3 & A & 0.0474   & 14:14:47.47 & $-$00:00:11.05        & 10.3  &5.\arcsec34      \\
SDSS J141447.15-000013.3 & B & 0.0475 & 14:14:47.14 & $-$00:00:13.10         & 10.2  &4.97 kpc  \\
\hline
SDSS J154403.67+044610.1 & A & 0.0416   & 15:44:03.66 & +04:46:10.11         & 10.5   &4.\arcsec19      \\
SDSS J154403.45+044607.5 & B & 0.0420  & 15:44:03.46 & +04:46:07.53         & 10.2  &3.48 kpc  \\
\hline

\end{tabular}
\caption{Sample of AGN pair candidates.
Column (1): Name of the galaxy. 
Column (2): Nucleus identifier. 
The two nuclei in each system are labeled “A” and “B”. 
Column (3): Redshift. Values are adopted from \cite{2025ApJS..281...25P}.
Columns (4) and (5): Right Ascension and Declination. 
Column (6): The stellar mass is derived from the 2MASS $Ks$-band image, adopting $\rm M/L=0.6$ \citep{2014AJ....148...77M} and measuring within a circular aperture of radius $3\arcsec.0$ centered on the nucleus.
Column (7): Projected separation between the nuclei in arcseconds and kiloparsecs. }
\label{tab:sample}
\end{table*}

\begin{figure*}[ht!]
\includegraphics[width=1\textwidth,trim=0 0 0 0]{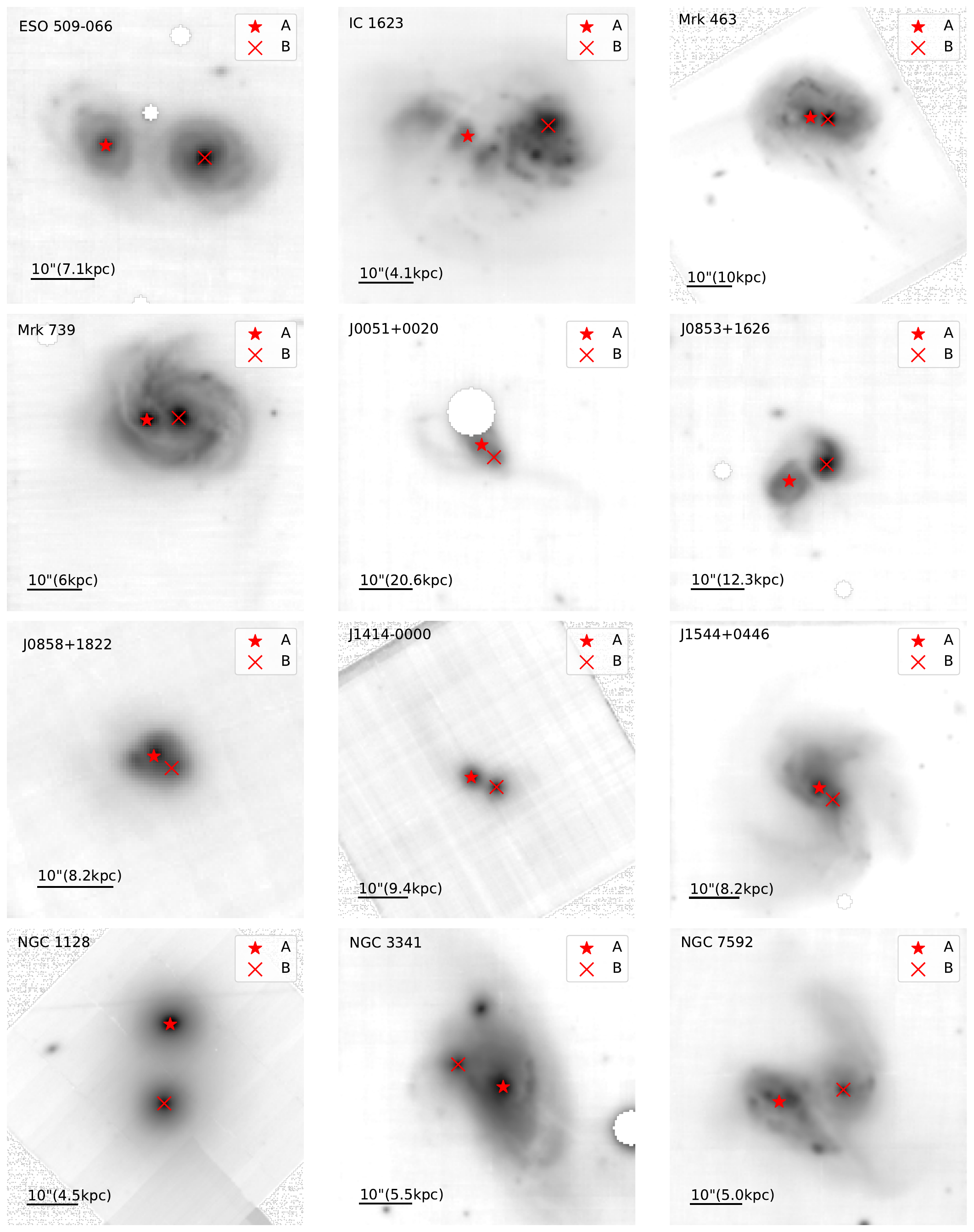}
\caption{MUSE continuum images of the AGN pair candidates. 
The images display the mean flux calculated over the rest-frame wavelength range of $5300$ to $5700\rm\,\AA$.
The positions of the nuclei are marked with red star and “X” symbols.
North is up and East is to the left.
Bright foreground stars, when present, are masked.}
\label{fig:MUSE continuum}
\end{figure*}

To select close AGN pairs suitable for detailed spatially resolved analysis, we performed a cross-match between the Big Multi-AGN Catalog \citep[The Big MAC;][]{2025ApJS..281...25P} and the public data archive of the VLT/MUSE. 
\cite{2025ApJS..281...25P} assembled The Big MAC by conducting a systematic, iterative keyword search of the NASA ADS database, reviewing $\sim600$ refereed articles published between 1970 and 2020 and compiling all confirmed and candidate multi-AGN systems across all selection methods, redshifts, and galaxy mass ratios. 
The resulting catalog contains a total of $>5700$ multi-AGN systems, including $\sim4300$ dual AGN/SMBH candidates (projected separations $\sim0.03$–-$110\rm\, kpc$), $\sim1300$ binary AGN/SMBH candidates, $\sim500$ recoiling AGN/SMBH candidates, and $\sim50$ N-tuple AGN systems.
As our primary scientific goal is to investigate the detailed kinematic structure and ionization properties of the gas within the interacting systems, we imposed some specific selection criteria on the cross-matched targets. 
First, we limited our sample to the local universe by requiring a redshift of $z\lesssim0.1$. 
This redshift cutoff ensures that we can achieve a physical spatial resolution sufficient to map the circumnuclear regions and feedback signatures. 
Second, we required a projected angular separation between the nuclei of $\Delta \theta \geq2.\arcsec0$. 
Given that the typical seeing-limited spatial resolution of MUSE in Wide Field Mode (WFM) is approximately $0.\arcsec6$--$1.\arcsec2$, this separation threshold guarantees that the two nuclei are spatially distinguishable and minimizes the contamination between their respective Point Spread Functions (PSFs). 
Third, to study the close AGN pairs, the projected distance is required to be $r_{\rm p}\leq 20\rm\,kpc$.
The final selected 12 AGN pair candidates are listed in Table~\ref{tab:sample}. 
The continuum images of these systems, obtained from the MUSE data, are presented in Figure~\ref{fig:MUSE continuum}.

Our sample is drawn from archival MUSE observations obtained under various observing programs. 
Rather than performing a rigorous statistical analysis of this sample, we focus on the detailed characterization of the individual systems.

\subsection{Observations and Data Reduction} \label{subsec:data}

\begin{table*}[]
\centering
\begin{tabular}{ccccc}
\hline
\hline
name        & Mode          & Exposure time & Program ID     & PI                        \\ \hline \hline
ESO 509-066 & MUSE-WFM & 1680s         & 0103.A-0637    & B. Husemann               \\ \hline
IC 1623     & MUSE-WFM-AO   & 3600s         & 0100.B-0116    & C. Carollo            \\ \hline
Mrk 463     & MUSE-WFM & 5800s         & 095.B-0482     & E. Treister               \\ \hline
Mrk 739     & MUSE-WFM & 5856s         & 095.B-0482     & E. Treister   \\ \hline
NGC 1128    & MUSE-WFM & 4800s         & 095.B-0482     & E. Treister               \\ \hline
NGC 3341    & MUSE-WFM & 4900s         & 110.23WR     & F. Bian               \\ \hline
NGC 7592    & MUSE-WFM & 2805s         & 0101.D-0748    & H. Kuncarayakti           \\ \hline
SDSS J0051+0020 & MUSE-WFM & 4200s         & 110.23WR       & F. Bian                   \\ \hline
SDSS J0853+1626 & MUSE-WFM & 8400s         & 110.23WR       & F. Bian                   \\ \hline
SDSS J0858+1822 & MUSE-WFM & 10500s        & 110.23WR       & F. Bian                   \\ \hline
SDSS J1414$-$0000 & MUSE-WFM & 11200s        & 109.238W       & F. Bian                   \\ \hline
SDSS J1544+0446 & MUSE-WFM & 10800s        & 109.238W       & F. Bian                   \\ \hline
\end{tabular}
\caption{Summary of the MUSE observation used in this paper.}
\label{tab:sum_of_obs}
\end{table*}

In this study, we utilize only WFM data from VLT/MUSE. 
A complete list of the observations used in this paper is provided in Table \ref{tab:sum_of_obs}.
The spaxel size of MUSE WFM data is 0.\arcsec2, but the spatial resolution is limited by the seeing ($\sim 0.\arcsec6$--$1\arcsec$).
The average spectral resolution is $R \sim 3000$, corresponding to a velocity dispersion of $\sim 50\rm\, km\,s^{-1}$.
We downloaded the pipeline processed data from ESO Archive Science Portal\footnote{$\rm https://archive.eso.org/scienceportal/home$}.

\begin{figure*}[ht!]
\includegraphics[width=1\textwidth,trim=0 0 0 0]{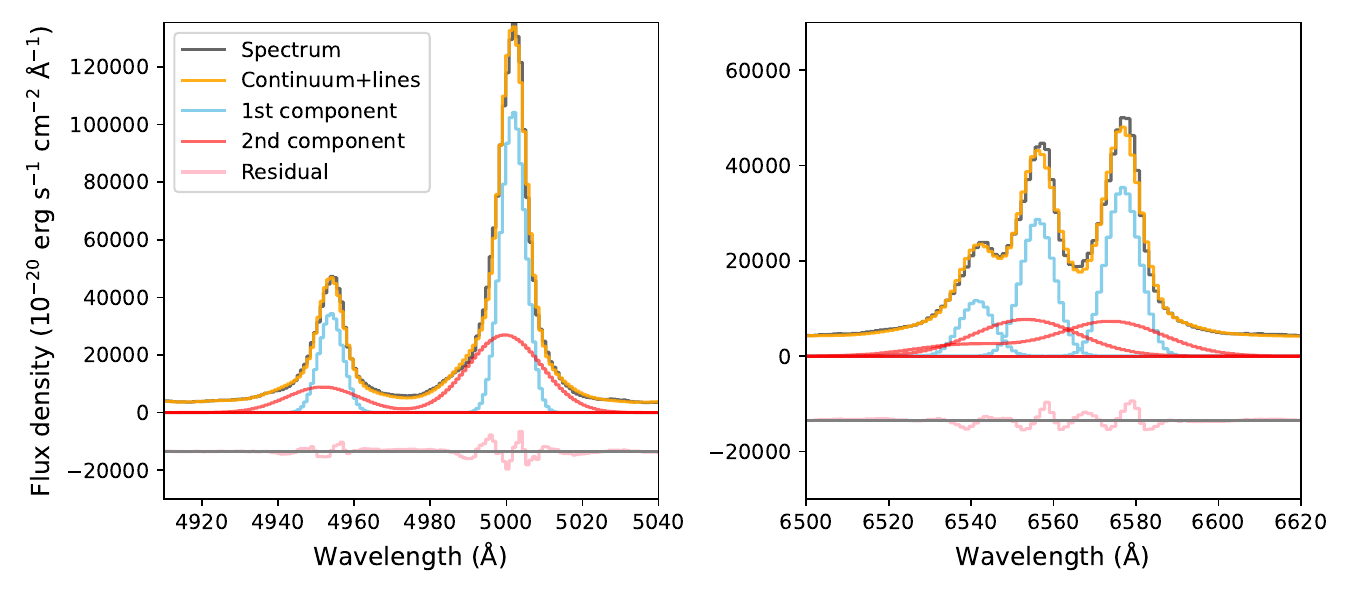}
\caption{Spectral analysis example showing nucleus B of ESO 509-066 with zoomed-in views of the rest-frame wavelength ranges $4910$--$5040\rm\,\AA$ (left) and $6500$--$6620\rm\,\AA$ (right).
The black line represents the observed data, and the orange line shows the total best-fit model (continuum plus emission lines). 
The light blue and red lines indicate the first (narrow) and second (broad) Gaussian components, respectively. 
The residuals are shown in pink and are offset below zero for clarity. 
}
\label{fig:example_spec}
\end{figure*}

To enhance the signal-to-noise ratio (S/N) for faint extended structures, we spatially rebinned the data by a factor of two. This results in a spaxel size of 0.\arcsec4. 
We used the Penalized Pixel-Fitting code \citep[pPXF;][]{2004PASP..116..138C,2023MNRAS.526.3273C} to fit the spectrum of each spaxel within the rest-frame wavelength range of $\rm 4800$--$6900\rm\,\AA$, to cover the main optical emission lines.
We modeled the stellar continuum using templates from the Flexible Stellar Population Synthesis (FSPS) library \citep{2009ApJ...699..486C,2010ApJ...712..833C}.
When fitting the continuum, the main emission lines are masked with a masking width of $1000\rm\, km\,s^{-1}$.
After determining the best-fit stellar continuum, we modeled the major gas emission lines using two Gaussian components. 
The fitted lines included H$\beta$, [O~{\sc{iii}}]$\lambda\lambda$4959,5007, [O~{\sc{i}}]$\lambda$6300, H$\alpha$, [N~{\sc{ii}}]$\lambda\lambda$6548,6583, and [S~{\sc{ii}}]$\lambda\lambda$6716,6731.
For each Gaussian component, we assumed the velocity ($v$) and velocity dispersion ($\sigma$) were identical across all emission lines.
We imposed a lower limit of $\sigma_{1} \geq 50\rm\, km\,s^{-1}$ for the first gas component.
We note that this lower limit corresponds to the instrumental spectral resolution of MUSE ($R \sim 3000$, or $\sigma_{\rm inst}\sim 50\rm\, km\,s^{-1}$).
As a result, the intrinsic velocity dispersion of the first component ($\sigma_{\rm 1st}$) remains unresolved in spaxels where $\sigma_{\rm 1st}$ reaches this boundary, and the reported $\sigma_{\rm 1st}$ values in those cases should be regarded as upper limits to the true gas velocity dispersion.
Additionally, we required the second component to be broader than the first by least $50\rm\,km\,s^{-1}$ ($\sigma_{2}-\sigma_{1}\geq50\rm\,km\,s^{-1}$). 
Figure \ref{fig:example_spec} shows an example spectrum of SDSS J085312.70+162615.5 (hereafter J0853+1626).
To ensure robust detections and exclude spurious signals in isolated spaxels, we retained only components with $\rm S/N \geq 5$ and line flux $\geq 5 \times 10^{-19}\rm\, erg\, s^{-1}\, cm^{-2}$.

Two nuclei in our sample have been classified as Type 1 AGNs: ESO 509-066A \citep{2017ApJ...850..168K} and Mrk 739A \citep{1987A&A...171...41N,2011ApJ...735L..42K,2021ApJ...911..100T}. 
For these sources, we included an additional broad Gaussian component to model the emission from the AGN broad-line region (BLR). 
We applied a three-step procedure to handle this broad emission. 
First, we fitted the spectrum of the brightest spaxel using three gas components (first, second, and broad components). 
This allowed us to determine the velocity and velocity dispersion of the broad component. 
Second, we fitted the spectra of all spaxels within the central region ($\leq1.\arcsec6$) with the same gas components.
During this step, we fixed the kinematics of the broad component to the values obtained from the brightest spaxel. 
Third, we subtracted the modeled broad component from the spectra in this central region.
This step allowed us to isolate the narrower emission lines (first and second components) for further analysis.

We corrected the emission line fluxes for dust extinction using the Balmer decrement (the flux ratio of H$\alpha$ to H$\beta$). 
We adopted the attenuation law of \cite{2000ApJ...533..682C} with $\rm R_{V}=3.1$.
Additionally, we assumed an intrinsic Balmer decrement of $\rm H\alpha / H\beta=2.86$ \citep[][]{2006agna.book.....O}.

\section{Results} \label{sec:result}

Here, we will present the spatially resolved results of our 12 pairs, including the surface brightness, kinematics, and BPT maps of each component.

\subsection{ESO 509-066} \label{subsec:ESO 509-066}

\begin{figure*}[ht!]
\includegraphics[width=1\textwidth,trim=0 0 0 0]{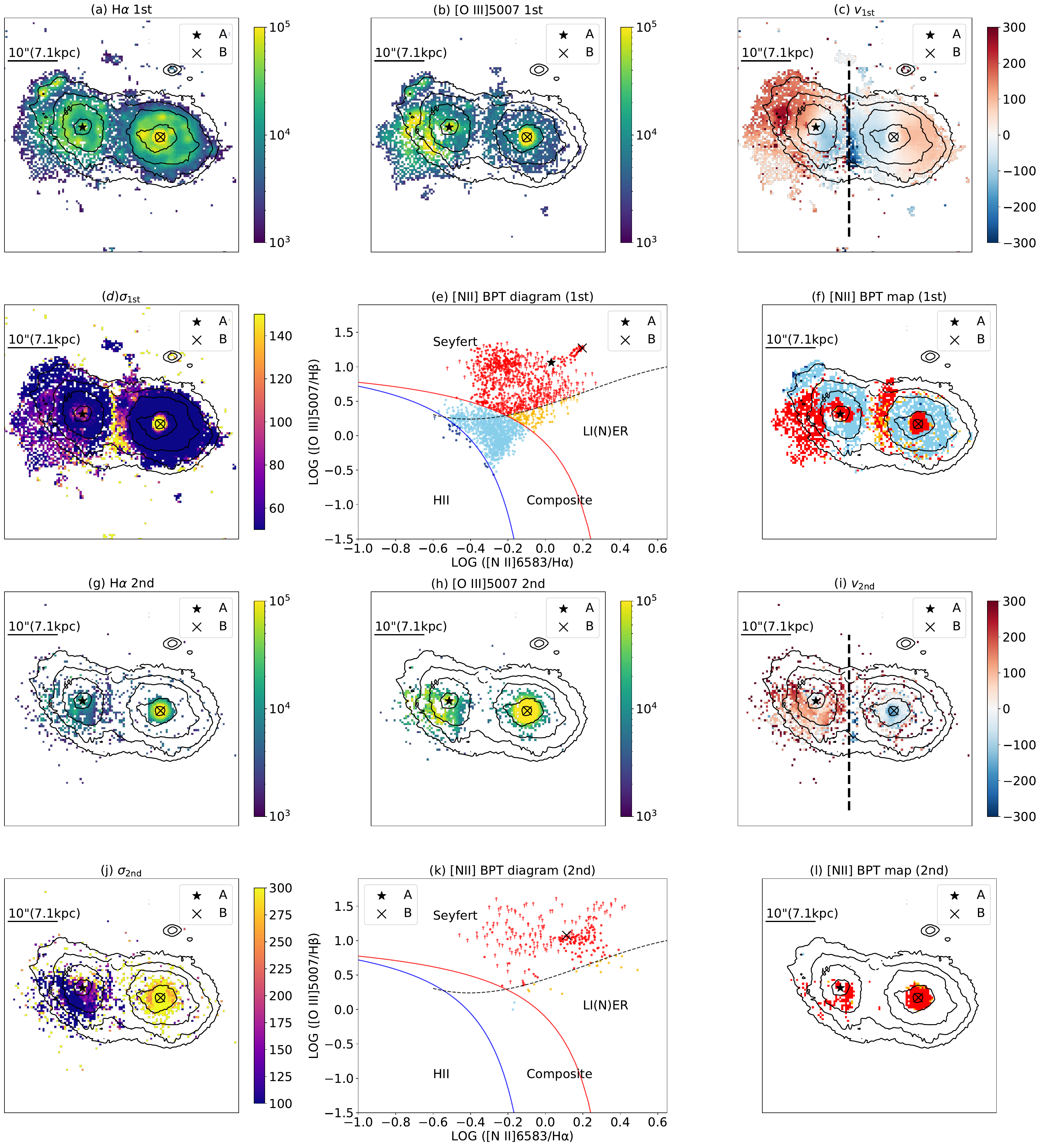}
\caption{Surface brightness, kinematics, and BPT maps of ESO 509-066. 
(a) and (b): the surface brightness map for the first gas component of H$\alpha$ and [O~{\sc{iii}}]$\lambda5007$, respectively.
(c) and (d): the velocity and velocity dispersion maps for the first component, respectively.
(e) and (f): the BPT diagnostic diagram and map for the first component, respectively.
The blue solid line \citep{2003MNRAS.346.1055K} and red solid line \citep{2001ApJ...556..121K} separate the H~{\sc ii} and composite regions. 
The black dashed line \citep{2025arXiv250517843C} separates the Seyfert and LI(N)ER regions.
(g) -- (l) are similar to (a) -- (f) but for the second gas component.
To better illustrate the rotating disk structures associated with each nucleus, the velocity maps (panels (c) and (i)) adopt different velocity references on either side of the dashed line: velocities to the left are shown relative to the systemic velocity of the first component of nucleus A  ($-590\rm\,km\,s^{-1}$), while velocities to the right are shown relative to the systemic velocity of the first component of nucleus B ($-265\rm\,km\,s^{-1}$).
Star and ``X” symbols mark the locations of Nuclei A and B, respectively. 
We display only spaxels with S/N$\ge5$ and a surface brightness $\ge 3.1\times10^{-18}\rm\,erg\,s^{-1}\,cm^{-2}\,arcsec^{-2}$.
Black contours show the stellar continuum from Figure.~\ref{fig:MUSE continuum} with signal-to-noise ratios (S/N) of 3, 5, 10, 20, and 50. 
}
\label{fig:E509_flux_kinematics_bpt}
\end{figure*}

ESO 509-066 is a galaxy pair located at $z = 0.034$ with a projected nuclear separation of $\sim 11\rm\, kpc$. 
Previous XMM-Newton observations of this system revealed evidence for two moderately obscured ($N_{\rm H}\sim10^{22}\rm\,cm^{-2}$), X-ray luminous nuclei that are spatially coincident with the optical nuclei with $L_{\rm2-10keV,A}=3.3\times10^{43}\rm\,erg\,s^{-1}$ and $L_{\rm2-10keV,B}=1.4\times10^{43}\rm\,erg\,s^{-1}$ \citep{2005A&A...429L...9G}. 
Subsequently, \cite{2017ApJ...850..168K} confirmed this system as a dual AGN pair using {\it Chandra} and Swift/XRT observations. 
They also detected a broad H$\alpha$ emission line in nucleus A using Keck/LRIS optical spectra, which is also presented in the MUSE data.

In this work, we attempt a first-order classification of the ionized-gas components in our sample using both kinematics and morphology, with the goal of separating rotating disks, tidal structures, and gas outflows. 
Here, we focus on these three broad categories as a practical framework rather than aiming for a fully exhaustive decomposition. 
Beyond them, merging systems can host additional, physically distinct gas components (e.g., inflows, shocked/interaction-interface gas, and mixed components arising from line-of-sight superposition). 
In this paper, we do not attempt to separately identify and interpret those additional components.
For extended tidal structures (e.g., tails or shells), we require (i) relatively low velocity dispersion, $\sigma<100\rm\,km\,s^{-1}$, and (ii) a velocity field that remains rotation-dominated. 
Morphologically, these components are identified as tails or shell-like features.
Because the velocity dispersion of the second gas component is always $\geq100\rm\,km\,s^{-1}$, this component is not classified as a tidal structure.
For outflows, we require (i) elevated velocity dispersion, $\sigma>150\rm\,km\,s^{-1}$, and (ii) velocity offsets that are inconsistent with purely rotating disk gas. 
Observationally, outflowing gas typically exhibit a blueward shift relative to the systemic velocity, or show paired blue- and redshifted components that can trace two-sided or biconical outflow geometries when projected onto the sky \citep[e.g.][]{2019A&A...622A.146M,2013ApJ...768...75R,2020AJ....159..167L}.
If a gas component cannot be classified as a rotating disk, tidal structure, or gas outflow, we classify it as ambiguous.

Figure \ref{fig:E509_flux_kinematics_bpt} displays the spatially resolved surface brightness and kinematic maps of ESO 509-066. 
In this work, we use ESO 509-066A and nucleus A interchangeably to refer to the same nucleus in ESO 509-066. 
The same convention is adopted for the other galaxies.
For the first gas component, we observe galactic-scale extended ionized gas around both nuclei, traced by H$\alpha$ and [O~{\sc iii}]. 
To better present the rotating disk structures associated with each nucleus, the velocity maps (panels (c) and (i)) adopt different velocity references on either side of the dashed line. 
Velocities to the left of the dashed line are shown relative to the systemic velocity of the first component of nucleus A ($-590\rm\,km\,s^{-1}$), while those to the right are presented relative to the systemic velocity of the first component of nucleus B ($-265\rm\,km\,s^{-1}$).
This indicates the presence of ionized gas disks in both systems. 
Diffuse gas with low velocity dispersion extends beyond the stellar disk of ESO 509-066A to the east. 
The $v_{\rm 1st}$ map clearly shows rotational velocity patterns in both galaxies. 
This feature may be associated with tidal stripping.
We also detect extended features in the second gas component around both nuclei. 
In the $v_{\rm 2nd}$ map, spaxels surrounding nucleus A consistently show redshifted velocity ($v_{\rm 2nd}\sim50$--$200\rm\,km\,s^{-1}$) and mild velocity dispersion ($\sigma_{\rm 2nd}\sim100$--$200\rm\,km\,s^{-1}$).
In contrast, the nuclear region of nucleus B exhibits blueshifted $v_{\rm 2nd}$ and higher $\sigma_{\rm 2nd}\sim320$--$550\rm\,km\,s^{-1}$. 
These properties suggest that the second gas component of nucleus B originates from outflowing gas. 
Conversely, the second gas component around nucleus A does not appear to be associated with an outflow.
The classification of the extended ionized gas structures is summarized in Table~\ref{tab:gas_structure}.

Figure \ref{fig:E509_flux_kinematics_bpt} presents the BPT diagnostic diagrams and maps for ESO 509-066. 
For spaxels where H$\beta$ is undetected, we adopt an upper limit for the H$\beta$ flux of $F_{\rm H\beta}=F_{\rm H\alpha}/3$. 
For the first gas component, both nucleus A and nucleus B fall within the Seyfert region in the BPT diagram. 
This result confirms that both nuclei host AGNs. 
A ring-like structure surrounding nucleus B is classified as a composite region in the BPT diagram, suggesting that SF may play a significant role in this area. 
Consistent with this interpretation, the H$\alpha$ map (Figure~\ref{fig:E509_flux_kinematics_bpt}(a)) reveals clumpy H$\alpha$ emission within the ring-like structure, further supporting the presence of a SF ring.
Regarding the second gas component, nucleus B is classified as a Seyfert source in the BPT diagram, and the circumnuclear region of nucleus A also falls within the Seyfert region. 
The BPT classification of ESO 509-066 and other targets are presented in Table~\ref{tab:bpt table}.

\begin{table}[ht]
\centering
\begin{tabular}{ccc}
\hline
\hline
name        & \multicolumn{2}{c}{Gas structure}   \\ 
            & 1st Component      & 2nd Component    \\ \hline \hline
ESO 509-066A & disk+tails       & ambiguous          \\ 
ESO 509-066B & disk+tails       & outflow     \\ \hline
IC1623A     & disk+tails         & outflow           \\ 
IC1623B     & disk+tails         & outflow           \\ \hline
Mrk463A     & tails              & outflow              \\ 
Mrk463B     & tails              & outflow             \\ \hline
Mrk739A     & tails         & outflow              \\ 
Mrk739B     & disk+tails         & ambiguous            \\ \hline
NGC 1128A    & compact            & compact outflow?     \\ 
NGC 1128B    & compact            & compact outflow?        \\ \hline
NGC 3341A    & disk+tails         & outflow?       \\ 
NGC 3341B    & tails              & outflow                \\ \hline
NGC 7592A    & disk+tails         & ambiguous          \\ 
NGC 7592B    & disk+tails         & outflow?           \\ \hline
J0051+0020A & disk?              & compact                \\ 
J0051+0020B & disk?              & compact               \\ \hline
J0853+1626A & disk+tails         & ambiguous           \\ 
J0853+1626B & disk+tails         & outflow?         \\ \hline
J0858+1822A & disk+tails         & outflow           \\ 
J0858+1822B & ambiguous         & outflow         \\ \hline
J1414$-$0000A & disk               & compact outflow        \\ 
J1414$-$0000B & disk               & compact outflow?       \\ \hline
J1544+0446A & disk+tails         & outflow             \\ 
J1544+0446B & tails              & outflow              \\ \hline
\end{tabular}
\caption{Gas structures of the extended ionized gas.
The gas structures are classified as rotating disks, tidal tails, outflows, or ambiguous (i.e., not attributable to any of the three categories). 
We classify nearly unresolved structures as compact. }
\label{tab:gas_structure}
\end{table}

\subsection{IC 1623} \label{subsec:IC 1623}

\begin{table}[]
\centering
\begin{tabular}{ccc}
\hline
\hline
name        & \multicolumn{2}{c}{BPT type}  \\ 
            & 1st Component      & 2nd Component \\ 
\hline
\hline
ESO509-066A & AGN      & AGN             \\ 
ESO509-066B & AGN      & AGN            \\ \hline
IC1623A     & AGN      & Composite              \\ 
IC1623B     & Composite    & Composite           \\ \hline
Mrk463A     & AGN      & AGN           \\ 
Mrk463B     & AGN  & AGN         \\ \hline
Mrk739A     & AGN      & SF              \\ 
Mrk739B     & Composite    & AGN           \\ \hline
NGC1128A    & LI(N)ER  & -              \\ 
NGC1128B    & LI(N)ER  & LI(N)ER      \\ \hline
NGC3341A    & Composite    & Composite         \\ 
NGC3341B    & AGN      & AGN           \\ \hline
NGC7592A    & Composite    & Composite      \\ 
NGC7592B    & Seyfert  & AGN          \\ \hline
J0051+0020A & Composite    & Composite           \\ 
J0051+0020B & AGN      & LI(N)ER           \\ \hline
J0853+1626A & AGN      & AGN          \\ 
J0853+1626B & Composite    & LI(N)ER         \\ \hline
J0858+1822A & AGN       & AGN          \\ 
J0858+1822B & LINER     & AGN            \\ \hline
J1414-0000A & Composite    & AGN            \\ 
J1414-0000B & AGN      & AGN            \\ \hline
J1544+0446A & LI(N)ER  & LI(N)ER      \\ 
J1544+0446B & LI(N)ER  & LI(N)ER      \\ \hline

\end{tabular}
\caption{Results of BPT diagnostic diagrams for all nuclei. 
The first column lists the nucleus name. 
The second and third column present the classifications of the first and the second gas component based on the [N~{\sc ii}] BPT diagram, respectively. }
\label{tab:bpt table}
\end{table}

IC 1623 (also known as VV 114 and Arp 236) is a nearby ($z = 0.020067$) mid-stage merger \citep{1992ApJS...83...29S}. 
It is classified as a luminous infrared galaxy (LIRG) with a total infrared luminosity of $L_{\rm IR,8-1000 \mu m} = 4.5\times10^{11}\,L_{\odot}$ \citep{2009PASP..121..559A}. 
Although \cite{2010RAA....10..309W} selected IC 1623 as a dual AGN candidate, they suggested that the X-ray emission is likely dominated by nuclear or circumnuclear starbursts rather than the AGN activity. 
This conclusion was based on the lack of Fe K line detections and the weak emission in the hard X-ray band.
Spectroscopic observations using the WiFeS optical IFS revealed two ionized gas components and elevated diagnostic line ratios across the merger \citep{2011ApJ...734...87R}. 
Recently, \cite{2025ApJ...988..230K} used MUSE data (program ID 097.B-0427; PI: G. Privon) to study the ionized gas in this system. 
They detected a relatively broad emission-line component ($\sigma \sim 100$–-$300\rm\, km\,s^{-1}$) extending across much of the galaxy. 
Based on the line ratios, kinematics, and energetics, they interpreted this component as a starburst-driven superwind. 
In this work, we analyze a MUSE dataset with a longer exposure time (program ID 0100.B-0116; PI: C.M. Carollo) to characterize the spatially resolved properties of the ionized gas.

The spatially resolved surface brightness and kinematics maps for IC 1623 are presented in Sect~\ref{sec:maps_others}.
The $v_{\rm 1st}$ map reveals a rotational velocity pattern around nucleus B. 
Most spaxels in this region exhibit low $\sigma_{\rm 1st}$, which suggests the presence of a rotating gas disk. 
For the second component, extended ionized gas is observed across nearly the entire merger. 
In the southern region, we detect blueshifted $v_{\rm 2nd}$ (up to $-250\rm\,km\,s^{-1}$) and relatively high $\sigma_{\rm 2nd}$ ($\sim200\rm\,km\,s^{-1}$), while the northern region exhibits redshifted $v_{\rm 2nd}$ (up to $220\rm\,km\,s^{-1}$). 
This kinematic structure indicates an ionized gas outflow. 
These kinematic results are consistent with the findings of \cite{2025ApJ...988..230K}.

Spatially resolved BPT diagrams and maps of IC 1623 are shown in Sect~\ref{sec:maps_others}. 
For the first gas component, IC 1623A is classified as a Seyfert nucleus in the BPT diagram. 
Previous observations with Keck and the James Webb Space Telescope (JWST) revealed that IC 1623A contains two cores. 
These cores are consistent with a dust-buried starburst and a heavily obscured AGN, respectively \citep{2001AJ....122.1213S,2022ApJ...940L...8E,2023ApJ...944L..50R,2024ApJ...966..166B}. 
Consequently, we suggest that the optical line ratios in IC 1623A originates from the combined SF and AGN activities.
IC 1623B is classified as a composite region in the BPT diagram and previous studies indicate that IC 1623B is dominated by starburst activity \citep[e.g.,][]{2011ApJ...734...87R,2022ApJ...940L...8E,2025ApJ...988..230K}. 
Therefore, IC 1623 might not be a true AGN pair.
For the second gas component, most spaxels fall within the composite region in the BPT diagram, while some spaxels at the edges are classified as Seyfert and LI(N)ER regions. 
This result suggests that the line ratios may be enhanced by the gas outflow and shocks in IC 1623, which is consistent with previous works \citep{2011ApJ...734...87R,2025ApJ...988..230K}.

\subsection{Mrk 463} \label{subsec:Mrk 463}

Mrk 463 is a late-stage merger and ultraluminous infrared galaxy (ULIRG) at $z=0.0508$ featuring two nuclei \citep{1991AJ....102.1241M}. 
\cite{2008MNRAS.386..105B} classified the system as a dual AGN based on {\it Chandra} X-ray observations. 
Mrk 463E (nucleus A) and Mrk 463W (nucleus B) have absorption-corrected luminosities of $L_{\rm2-10keV,A}=1.5\times10^{43}\rm\,erg\,s^{-1}$ and $L_{\rm2-10keV,B}=3.8\times10^{42}\rm\,erg\,s^{-1}$, respectively. 
Both nuclei are heavily obscured, with column densities of $N_{\rm H,A}=7.1\times10^{23}\rm\,cm^{-2}$ and $N_{\rm H,B}=3.2\times10^{23}\rm\,cm^{-2}$. 
Using VLT/MUSE observations, \cite{2018ApJ...854...83T} confirmed that both nuclei are classified as Seyferts in the BPT [N~{\sc ii}] and [S~{\sc ii}] diagrams. 
They also detected a biconical outflow with $v_{\rm out}>600\rm\,km\,s^{-1}$ associated with nucleus A. 
Subsequently, \cite{2022MNRAS.511.2105K} analyzed Mrk 463 as part of the Swift-BAT AGN Spectroscopic Survey (BASS) \citep{2017ApJ...850...74K}. 
Using the same MUSE data, they confirmed the presence of the ionized gas outflow based on a broader [O~{\sc iii}] component. 
In this work, we use this MUSE data to reproduce the spatially resolved surface brightness, kinematic, and BPT maps of Mrk 463.

The spatially resolved surface brightness and kinematics maps of Mrk 463 are presented in Sect~\ref{sec:maps_others}. 
While \cite{2018ApJ...854...83T} and \cite{2022MNRAS.511.2105K} focused primarily on [O~{\sc iii}] flux and kinematics, we present flux maps for H$\alpha$ and [O~{\sc iii}] emission lines, along with kinematic maps for two gas components. 
For the first gas component, we detect extended ionized gas around both nuclei and within the spiral arms or tidal tails. 
We do not observe an obvious rotational velocity pattern around either nucleus. 
However, relatively high $\sigma_{\rm 1st}$ ($\sim220\rm\,km\,s^{-1}$) is detected around nucleus A.
For the second gas component, extended ionized gas traced by H$\alpha$ and [O~{\sc iii}] emission lines is present throughout the system. 
Around nucleus A, we detect a north-south velocity gradient ($-250\rm\,km\,s^{-1}$ to $+300\rm\,km\,s^{-1}$) with high $\sigma_{\rm 2nd}$ (up to $>300\rm\,km\,s^{-1}$). 
This suggests an ionized gas outflow, which is consistent with the results of \cite{2018ApJ...854...83T}. 
Similarly, around nucleus B, we observe a north-south velocity gradient ($-150\rm\,km\,s^{-1}$ to $+250\rm\,km\,s^{-1}$) with high $\sigma_{\rm 2nd}$ ($>300\rm\,km\,s^{-1}$), which may also indicate an ionized gas outflow. 
These outflow features could originate from a single outflow with a large opening angle driven by nucleus A, or they could be produced by two distinct outflows from the two nuclei. 
Higher spatial resolution data are required to confirm these scenarios.

Spatially resolved BPT diagrams and maps of Mrk 463 are shown in Sect~\ref{sec:maps_others}. 
For the first component, both nuclei are classified as Seyfert in the BPT diagram, consistent with previous results \citep[][]{2018ApJ...854...83T}. 
Combining the BPT classification with the previous X-ray observation \citep[][]{2008MNRAS.386..105B}, Mrk 463 is therefore likely a genuine AGN pair.
For the second gas component, most spaxels in the galaxy are classified as Seyfert.

\subsection{Mrk 739} \label{subsec:Mrk 739}

Mrk 739 is a merging system located at $z=0.02985$ that features double nuclei. 
\cite{2011ApJ...735L..42K} classified the system as a dual AGN based on {\it Chandra} X-ray observations. 
Mrk 739A and Mrk 739B have absorption-corrected X-ray luminosities of $L_{\rm2-10keV,A}=1.1\times10^{43}\rm\,erg\,s^{-1}$ and $L_{\rm2-10keV,B}=1.0\times10^{42}\rm\,erg\,s^{-1}$, respectively \citep{2011ApJ...735L..42K}. 
In the optical band, Mrk 739A exhibits broad Balmer emission lines, consistent with a Type 1 AGN \citep{1987A&A...171...41N,2011ApJ...735L..42K,2021ApJ...911..100T}. 
Using VLT/MUSE data, \cite{2021ApJ...911..100T} conducted a comprehensive morphological and kinematic study of Mrk 739. 
They proposed that the system is in an early merger stage, where Mrk 739B is a foreground, young SF galaxy undergoing its first passage with a background elliptical companion, Mrk 739A.

The spatially resolved surface brightness and kinematics maps of Mrk 739 are presented in Sect~\ref{sec:maps_others}. 
We present surface brightness maps for the H$\alpha$, and [O~{\sc iii}] emission lines, along with kinematic maps for two gas components, to supplement the findings of \cite{2021ApJ...911..100T}. 
For the first gas component, we detect extended ionized gas across the galaxy. 
We observe a large spiral structure with redshifted $v_{\rm 1st}$ extending northward from the circumnuclear region of nucleus A. 
This feature is consistent with the results of \cite{2021ApJ...911..100T}. 
The $\sigma_{\rm 1st}$ map exhibits low values across nearly the entire galaxy, with the exception of nucleus A. 
For the second gas component, extended ionized gas is presented around both nuclei and within a cloud in the northeast tail. 
Around Mrk 739A, we detect blueshifted velocities (up to $-250\rm\,km\,s^{-1}$) with high $\sigma_{\rm 2nd}$ ($>300\rm\,km\,s^{-1}$), which may suggest an ionized gas outflow. 
We also observe extended gas with relatively low $\sigma_{\rm 2nd}$ ($\sim100\rm\,km\,s^{-1}$) around Mrk 739B, which does not resemble either an outflow or a tidal structure.

Spatially resolved BPT diagrams and maps of Mrk 739 are shown in Sect~\ref{sec:maps_others}. 
For the first component, Mrk 739A is classified as a Seyfert, while Mrk 739B is classified as a composite region in the BPT diagram. 
Given that Mrk 739B has been confirmed as an AGN by \cite{2011ApJ...735L..42K}, Mrk 739B might be affected by both SF and AGN activities. 
In the whole galaxy, most spaxels are classified as Seyfert regions in the BPT map, except a composite and SF region extends from Mrk 739B to the southwest, which is consistent with the findings of \cite{2021ApJ...911..100T}. 
For the second component, spaxels around Mrk 739A are classified as H~{\sc ii} regions, potentially indicating a nuclear SF. 
The tail-like structure is dominated by AGN photoionization.

We find that Mrk 739A may have been a fading AGN over the past several $10^{4}\rm\, yr$, as discussed in Section~\ref{subsec:fading AGN}.

\subsection{NGC 1128} \label{subsec:NGC 1128}

NGC 1128, also known as 3C 75, is a well-known double radio source located at the center of the nearby galaxy cluster Abell 400 (A400). 
Early radio observations by \cite{1985ApJ...294L..85O} first revealed the double radio jets of 3C 75. 
Subsequently, \cite{2006A&A...453..433H} detected two X-ray sources that are spatially coincident with the dual radio cores. 
These radio and X-ray observations confirm that this system is an AGN pair.

The surface brightness and kinematic maps of NGC 1128 are presented in Sect~\ref{sec:maps_others}. 
For the first component, we detect almost no extended ionized gas emission, and line emission is observed only in the nuclei. 
Similarly, for the second gas component, only the double nuclei exhibit line emission. 
This emission features high velocity dispersion, with $\sigma_{\rm 2nd}\sim600\rm\,km\,s^{-1}$ (Table \ref{tab:kinematics-flux table}). 
We confirmed these results by inspecting the spectra of the two nuclei. 
The high $\sigma_{\rm 2nd}$ is presented in both Balmer lines and forbidden lines. 
Therefore, this feature cannot originate from the broad-line regions (BLRs) of the AGNs. 
Given the presence of strong radio jets, the ionized gas with high $\sigma_{\rm 2nd}$ could be associated with radio jets and shocks.

Spatially resolved BPT diagrams and maps of NGC 1128 are shown in Sect~\ref{sec:maps_others}. 
For the first components, both the two nuclei are classified as LI(N)ERs in the BPT diagram, which also suggests the ionized gas in these two nuclei might be associated with the radio jets.

\subsection{NGC 3341} \label{subsec:NGC 3341}

NGC 3341 is a nearby ($z=0.027339$) interacting system comprising a giant disk galaxy and two dwarf companions located within its disk \citep[][]{2013MNRAS.435.2335B}. 
The two offset nuclei are located at projected distances of $5.1\rm\,kpc$ (9.\arcsec5 to the northeast; NGC 3341B) and $8.4\rm\,kpc$ (15.\arcsec6 to the north; NGC 3341C) from the primary nucleus NGC 3341A. 
These companions may be dwarf elliptical or the spheroidal merger remnants of low-mass spirals \citep[][]{2013MNRAS.435.2335B}. 
Multi-wavelength observations have confirmed that NGC 3341B hosts a Seyfert 2 AGN \citep{2008ApJ...683L.119B,2013MNRAS.435.2335B}. 
\cite{2013MNRAS.435.2335B} also found that NGC 3341A is dominated by SF. 
Furthermore, optical, radio, and X-ray data show no evidence of AGN activity in NGC 3341C. 
\cite{2011ApJ...737..101L} identified another galaxy, SDSS J104229.80+050500.2 (hereafter J1042+0505), located at a projected distance of $\sim80.3\rm\,kpc$ ($142.\arcsec3$) from NGC 3341, as an AGN host. 

The spatially resolved surface brightness and kinematics maps of NGC 3341 are presented in Sect~\ref{sec:maps_others}. 
For the first component, we detect extended ionized gas emission across the MUSE FoV. 
The [O~{\sc iii}] emission is much less extended than the H$\alpha$ emission and is concentrated primarily around NGC 3341B.
We observe a rotational velocity field with low $\sigma_{\rm 1st}$ around NGC 3341A, indicating the presence of a rotating disk. 
A large tidal tail connects NGC 3341A and NGC 3341C, exhibiting a smooth $v_{\rm 1st}$ gradient and low $\sigma_{\rm 1st}$. 
For the second gas component, extended emission is found primarily around NGC 3341A and NGC 3341B. 
In the $v_{\rm 2nd}$ map, we observe blueshifted velocities with high $\sigma_{\rm 2nd}\sim 200\rm\,km\,s^{-1}$ around NGC 3341B, suggesting an ionized gas outflow. 
Around NGC 3341A, we detect a blueshifted component with a more moderate velocity dispersion ($\sigma_{\rm 2nd}\sim 150\rm\,km\,s^{-1}$), which might hint at a moderate outflow. 

Spatially resolved BPT diagrams and maps of NGC 3341 are shown in Sect~\ref{sec:maps_others}. 
For the first component, NGC 3341B is classified as a Seyfert in the BPT diagram, whereas NGC 3341A and NGC 3341C are classified as composite regions. 
The ionized gas surrounding NGC 3341C may be influenced by tidal stripping and/or photoionization from NGC 3341B. 
We find no evidence of AGN activity associated with NGC 3341C, which is consistent with \cite{2013MNRAS.435.2335B}. 
The circumnuclear region of NGC 3341B is dominated by Seyfert ionization in the BPT diagram, while the spaxels surrounding NGC 3341A are dominated by SF regions. 
For the second component, NGC 3341B is also classified as Seyfert in the diagram, and NGC 3341A is classified as a composite region.
These results suggest only NGC 3341B host an accreting AGN, while NGC 3341A and NGC 3341C might be dominated by SF.

\subsection{NGC 7592} \label{subsec:NGC 7592}

NGC 7592, also known as VV 731, is an interacting system at $z=0.02444$.
\cite{2002ApJS..143...47D} classified the western nucleus (NGC 7592B) as a Seyfert 2 galaxy and the eastern nucleus (NGC 7592A) as a starburst.
Although NGC 7592 is a candidate AGN pair, NGC 7592A exhibits little emission in either the soft or hard X-ray bands. 
This nucleus may be heavily obscured or simply intrinsically weak in X-rays \citep{2010RAA....10..309W}. 
Using {\it Chandra} data, \cite{2018A&A...620A.140T} confirmed the presence of AGN activity in NGC 7592B with a hard X-ray luminosity of $L_{\rm2-10keV,B}=5.9\times10^{40}\rm\,erg\,s^{-1}$.

The spatially resolved surface brightness and kinematics maps of NGC 7592 are presented in Sect~\ref{sec:maps_others}. 
For the first component, we detect extended ionized gas emission across the FoV of the MUSE data. 
The [O~{\sc iii}] emission is less extended than the H$\alpha$ emission. 
We observe two rotational velocity patterns with low $\sigma_{\rm 1st}$ around NGC 7592A and NGC 7592B, indicating the presence of two rotating gas disks. 
For the second gas component, we detect extended emission around both nuclei, although it is more extended around NGC 7592A. 
The $v_{\rm 2nd}$ and $\sigma_{\rm 2nd}$ maps reveal complex kinematic fields around the two nuclei, indicating a complicated dynamical scenario. 
Around NGC 7592B, high $\sigma_{\rm 2nd}$ may indicate the gas outflow or inflow. 
Future detailed study of NGC 7592 is needed to resolve the origins of these features.

Spatially resolved BPT diagrams and maps of NGC 7592 are shown in Sect~\ref{sec:maps_others}. 
For the first component, NGC 7592B is classified as a Seyfert in the BPT diagram, whereas NGC 7592A is classified as a composite region. 
Most spaxels surrounding NGC 7592A are dominated by SF or composite regions in the BPT diagrams. 
A biconical region centered on NGC 7592B and extending from the northwest to the southeast is classified as a Seyfert region. 
Another biconical region oriented roughly east-west is dominated by composite regions. 
This result suggests that SF and AGN activities may coexist around NGC 7592B, which is consistent with \cite{1992AJ....103..743R}. 
For the second component, NGC 7592B is also classified as a Seyfert in the BPT diagram, and NGC 7592A is classified as a composite region. 
Similar to the first component, the spaxels surrounding NGC 7592A are dominated by composite regions. 
In the biconical Seyfert region of NGC 7592B, the second gas component is also dominated by AGN photoionization.
Additionally, this region is roughly corresponding to the high $\sigma_{\rm 2nd}$ region, which might indicate the enhanced velocity dispersion is associated with the AGN activities.

Similar to Mrk 739A, we find that NGC 7592B may also have been a fading AGN over the past several $10^{4}\rm\, yr$, as discussed in Section~\ref{subsec:fading AGN}.

\subsection{J0051+0020} \label{subsec:J0051+0020}

SDSS J005114.10+002049.5 (hereafter J0051+0020) is a late-stage merger at $z=0.1126$ featuring two nuclei and two long tidal tails (see Figure \ref{fig:MUSE continuum}). 
This system was identified as a radio AGN pair based on Very Large Array (VLA) observations \citep{2015ApJ...799...72F,2015ApJ...815L...6F}. 
Subsequently, \cite{2019ApJ...883...50G} confirmed J0051+0020 as a dual AGN system using {\it Chandra} X-ray observations. 
They reported unabsorbed X-ray luminosities of $L_{\rm2-10keV,A}=10^{41.0}\rm\,erg\,s^{-1}$ and $L_{\rm2-10keV,B}=10^{41.5}\rm\,erg\,s^{-1}$ for nucleus A and nucleus B, respectively.

The spatially resolved surface brightness and kinematics maps of J0051+0020 are presented in Sect~\ref{sec:maps_others}. 
For the first gas component, we detect extended ionized gas around both nuclei. 
The H$\alpha$ emission exhibit a more extended morphology than the [O~{\sc iii}] emission. 
In the $v_{\rm 1st}$ map, although there is a bright foreground star affects the detection of extended ionized gas around J0051+0020A, there seems to be a rotational velocity feature around J0051+0020A, which indicates a gas disk.
For the second component, compact ionized gas (traced mainly by H$\alpha$) is presented in both nuclei. 

Spatially resolved BPT diagrams and maps of J0051+0020 are shown in Sect~\ref{sec:maps_others}. 
For the first component, J0051+0020A is classified as a composite region in the BPT diagram. 
This suggests that the dominant optical ionization source in J0051+0020A is not the AGN. 
Given that J0051+0020A is identified as an AGN in the radio and X-ray bands \citep{2015ApJ...799...72F,2015ApJ...815L...6F,2019ApJ...883...50G}, we suggest that both SF and AGN activity coexist in this nucleus. 
J0051+0020B is classified as a Seyfert in the BPT diagram, which is consistent with the radio and X-ray results \citep{2015ApJ...799...72F,2015ApJ...815L...6F,2019ApJ...883...50G}. 
For the second gas component, J0051+0020B is located in the LI(N)ER region, while J0051+0020A is classified as composite region.

\subsection{J0853+1626} \label{subsec:J0853+1626}

SDSS J085312.70+162615.5 (hereafter J0853+1626) is an interacting galaxy pair with a projected nuclear separation of $\sim 7.9\rm\, kpc$. 
\cite{2011ApJ...737..101L} classified J0853+1626 as an AGN pair using the BPT [N~{\sc ii}] diagram. 
They based this classification on emission line measurements from the MPA-JHU catalog \citep{2004MNRAS.351.1151B}.
\cite{2018MNRAS.478.1326L} detected hard X-ray emission from nucleus A with an unabsorbed luminosity of $L_{\rm 2-10\,keV}=2.2\times10^{42}\rm\, erg\,s^{-1}$. 
This indicates that nucleus A hosts an AGN. 
In contrast, only a soft X-ray source was detected at nucleus B. 
However, \cite{2018MNRAS.478.1326L} noted that the presence of a faint AGN in nucleus B could not be ruled out.

The spatially resolved surface brightness and kinematics maps of J0853+1626 are presented in Sect~\ref{sec:maps_others}. 
For the first gas component, the $v_{\rm 1st}$ map reveals a rotating gas disk around J0853+1626A. 
J0853+1626B also displays a velocity gradient from east to west, which suggests the presence of a gas disk around this nucleus as well. 
The velocity dispersion ($\sigma_{\rm 1st}$) is higher around nucleus B compared to nucleus A, which may imply that the disk around nucleus B could be more strongly disturbed.
The smooth distribution of flux and kinematics between the two nuclei indicates that the ionized gas in this merging system is mixing. 
Some extended tail-like structures with low $\sigma_{\rm 1st}$ can be observed in the outskirt region of this galaxy.
For the second gas component, extended features can also be detected around both J0853+1626A and J0853+1626B. 
Tail or shell-like structures are visible, particularly in the H$\alpha$ flux map. 
One feature is located northeast of nucleus A, and another lies to the southeast.
In the $v_{\rm 2nd}$ map, the northeast feature shows blueshifted velocities, while the southeast feature is redshifted. 
The $\sigma_{\rm 2nd}$ map shows relatively low velocity dispersion ($\sigma_{\rm 2nd}\sim 120\rm\,km\,s^{-1}$) in these regions. 
We detect no significant outflowing gas signatures around nucleus A, as the second gas component lacks high velocity dispersion. 
There appears to be another tail-like structure in the south region of J0853+1626B. 
Around J0853+1626B itself, we observe blueshifted $v_{\rm 2nd}$ and relatively high $\sigma_{\rm 2nd}$ values, which may indicate a gas outflow.

Spatially resolved BPT diagrams and maps of J0853+1626 are shown in Sect~\ref{sec:maps_others}. 
In the first gas component, the BPT map shows a Seyfert region extends northwest from J0853+1626A and connects to the circumnuclear region of J0853+1626B.  
A shell-like feature located southeast of nucleus A is classified as the composite region in the BPT diagram, which might indicate that SF occurs within this structure. 
J0853+1626B and the spaxels around J0853+1626B are primarily classified as composite regions, which suggests that nucleus B is dominated by SF activity, which is consistent with the result of \cite{2018MNRAS.478.1326L}.
For the second gas component, most spaxels around J0853+1626A fall within the Seyfert region in all diagnostic diagrams. 
In contrast, the spaxels around nucleus B are primarily classified as composite regions. 
Based on these BPT results, J0853+1626 is an AGN+SF pair rather than a dual AGN pair.

\subsection{J0858+1822} \label{subsec:J0858+1822}

SDSS J085837.53+182221.6 (hereafter J0858+1822) is a late-stage merging system at $z\sim0.059$. 
\cite{2011ApJ...737..101L} originally classified J0858+1822 as an AGN pair. 
However, \cite{2020A&A...639A.117H} argued that J0858+1822B is likely dominated by SF. 
Their conclusion was based on BPT [N~{\sc ii}] diagnostics using data from the Multi-Object Double Spectrograph (MODS) on the Large Binocular Telescope (LBT). 
They also reported that the peak of the [O~{\sc iii}] emission is located between the two continuum nuclei. 
Using {\it Chandra} observations, \cite{2020ApJ...900...79H} detected soft X-ray emission in both nuclei of J0858+1822, but no hard X-ray emission was observed. 
Furthermore, \cite{2020ApJ...900...79H} detected a soft X-ray source between the two nuclei. 
This feature is spatially consistent with the [O~{\sc iii}] result reported by \cite{2020A&A...639A.117H}.

The spatially resolved surface brightness and kinematics maps of J0858+1822 are presented in Sect~\ref{sec:maps_others}. 
For the first gas component, we detect extended ionized gas around both nuclei A and B. 
The H$\alpha$ emission exhibits a more extended morphology than the [O~{\sc iii}] emission. 
Notably, the peak of the [O~{\sc iii}] emission is located between the two continuum nuclei. 
This location coincides with the spaxels exhibiting the highest $\sigma_{\rm 1st}$ and is consistent with the findings of \cite{2020A&A...639A.117H}. 
The $v_{\rm 1st}$ map reveals a complex velocity field. 
For the second gas component, we observe extended ionized gas traced by H$\alpha$ and [O~{\sc iii}] emission lines. 
Similar to the first component, the peak of the [O~{\sc iii}] emission for the second component is located between the two continuum nuclei. 
In the $v_{\rm 2nd}$ and $\sigma_{\rm 2nd}$ maps, we observe redshifted velocities (up to $\sim200\rm\,km\,s^{-1}$) with high velocity dispersion (up to $\sim300\rm\,km\,s^{-1}$) to the southeast. 
Conversely, we detect blueshifted velocities (up to $\sim150\rm\,km\,s^{-1}$) with high velocity dispersion (up to $\sim250\rm\,km\,s^{-1}$) to the northwest. 
These kinematic features are consistent with a biconical gas outflow. 
The region with the highest $\sigma_{\rm 2nd}$ coincides with the peak of the second [O~{\sc iii}] component, which aligns spatially with the results from the first component. 

Spatially resolved BPT diagrams and maps of J0858+1822 are shown in Sect~\ref{sec:maps_others}. 
For the first component, the Seyfert region presents a biconical feature in the BPT map, and the center is consistent with the location of the peak of [O~{\sc iii}] emission and also the highest $\sigma_{\rm 1st}$.
J0858+1822A is classified as a Seyfert in the diagram, and J0858+1822B is classified as a LI(N)ER. 
For the second gas component, most spaxels are classified as Seyfert regions, while some edge spaxels are classified as LI(N)ERs. 

\subsection{J1414$-$0000} \label{subsec:J1414$-$0000}

SDSS J141447.15$-$000013.3 (hereafter J1414$-$0000) is a late-stage merging system at $z\sim0.0475$. 
\cite{2011ApJ...737..101L} originally classified J1414$-$0000 as an AGN pair. 
Subsequently, \cite{2020ApJ...900...79H} confirmed that both nuclei host AGNs based on {\it Chandra} X-ray observations. 
They reported luminosities of $L_{\rm2-10keV,A}=10^{41.73}\rm\,erg\,s^{-1}$ and $L_{\rm2-10keV,B}=10^{42.27}\rm\,erg\,s^{-1}$ for J1414$-$0000A and J1414$-$0000B, respectively.

The spatially resolved surface brightness and kinematics maps of J1414$-$0000 are presented in Sect~\ref{sec:maps_others}. 
For the first gas component, we detect extended ionized gas around both nuclei A and B. 
The H$\alpha$ emission exhibit a more extended morphology than the [O~{\sc iii}] emission. 
Around J1414$-$0000A, we observe a rotational velocity pattern accompanied by low $\sigma_{\rm 1st}$, which suggests the presence of a gas disk. 
Similarly, around nucleus B, we find low $v_{\rm 1st}$ with a gradient from northeast to southwest and low $\sigma_{\rm 1st}$, which also suggests the existence of a gas disk. 
For the second component, we observe compact ionized gas traced by H$\alpha$ and [O~{\sc iii}] emission lines around both nuclei.  
In nucleus A, the blueshifted $v_{\rm 2nd}$ and high $\sigma_{\rm 2nd}$ may indicate a compact ionized gas outflow. 
For nucleus B, the relatively high $\sigma_{\rm 2nd}$ may also indicate an outflow or perturbed gas.

Spatially resolved BPT diagrams and maps of J1414$-$0000 are shown in Sect~\ref{sec:maps_others}. 
For the first component, J1414$-$0000A is classified as a composite region in the BPT diagram. 
In contrast, J1414$-$0000B is classified as a Seyfert in the diagram. 
Based on the BPT diagnostics, the optical emission of J1414$-$0000A appears to be dominated by SF. 
However, hard X-ray detection indicates the presence of an AGN \citep{2020ApJ...900...79H}. 
For the second gas component, both J1414$-$0000A and J1414$-$0000B are classified as Seyfert, which indicates the second gas components of these two nuclei could be associated with the AGN activities.

\subsection{J1544+0446} \label{subsec:J1544+0446}

SDSS J154403.45+044607.5 (hereafter J1544+0446) is a late-stage merger at $z\approx0.042$ featuring two bright nuclei and tidal features. 
\cite{2011ApJ...737..101L} originally identified J1544+0446 as a candidate AGN pair. 
Subsequently, \cite{2019ApJ...882...41H} confirmed the system as a dual AGN using {\it Chandra} ACIS-S X-ray imaging spectroscopy.

The spatially resolved surface brightness and kinematics maps of J1544+0446 are presented in Sect~\ref{sec:maps_others}. 
For the first gas component, we detect extended ionized gas around both nuclei and within the spiral arm. 
We observe a rotational velocity pattern accompanied by low $\sigma_{\rm 1st}$ around J1544+0446A, which suggests the presence of a gas disk. 
In contrast, no obvious rotating disk features are visible around J1544+0446B. 
For the second gas component, we find extended ionized gas traced by H$\alpha$ and [O~{\sc iii}] emission lines around both nuclei. 
Around J1544+0446A, we detect a velocity gradient along the northeast-southwest direction with mild $\sigma_{\rm 2nd}$ (up to $\sim150\rm\,km\,s^{-1}$), which may indicate an ionized gas outflow. 
Similarly, around nucleus B, we observe a velocity gradient (ranging from $-150\rm\,km\,s^{-1}$ to $+100\rm\,km\,s^{-1}$) with high $\sigma_{\rm 2nd}$ (up to $\sim350\rm\,km\,s^{-1}$) along the northwest-southeast direction. 
This kinematic signature also suggests an ionized gas outflow.

Spatially resolved BPT diagrams and maps of J1544+0446 are shown in Sect~\ref{sec:maps_others}. 
For the first component, J1544+0446A is classified as Seyfert in the BPT diagram, and J1544+0446B is classified as a LI(N)ER.  
We note that the ionized gas cloud located to the southeast ($\sim15\rm\,kpc$ from the nuclei) is dominated by AGN photoionization in the BPT diagram. 
For the second gas component, both J1544+0446A and J1544+0446B are classified as LI(N)ERs in the BPT diagrams. 
The extended ionized gas surrounding these two nuclei is also dominated by LI(N)ER emission.

Similar to Mrk 739A and NGC 7592B, J1544+0446A may also have been a fading AGN over the past several $10^{4}\rm\, yr$, as discussed in Section~\ref{subsec:fading AGN}.

\section{Discussion} \label{sec:discussion}

\subsection{The nature of the AGN pair candidates} \label{subsec:nature of AGN-pair}

\begin{figure*}[ht!]
\includegraphics[width=1\textwidth,trim=0 0 0 0]{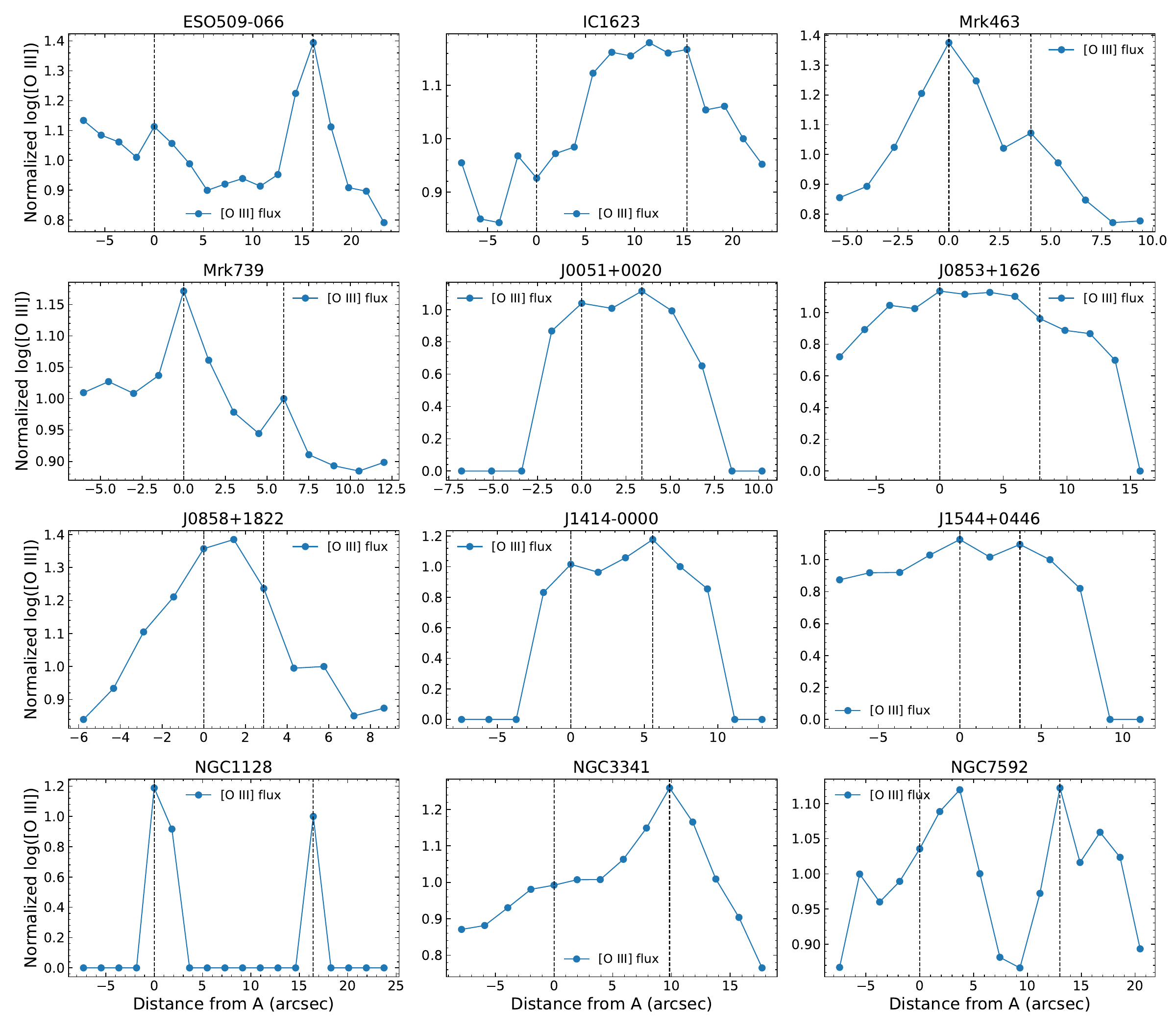}
\caption{Normalized [O~{\sc iii}] surface brightness along the two nuclei (from nucleus A to B). 
The surface brightnesses are extracted from a pseudo-slit with a 
$3\arcsec$ width along the two nuclei in the [O~{\sc iii}] surface-brightness maps, and are normalized by the median surface brightness of [O~{\sc iii}]. 
The locations of the two nuclei are indicated by the two vertical dashed lines, with nucleus A fixed at 0.}
\label{fig:o3_flux_AB_profile}
\end{figure*}

To assess whether the ionization state of one nucleus could be affected by radiation from the companion AGN, we extract the [O~{\sc iii}] surface brightness profiles along the axis connecting nuclei A and B (Figure~\ref{fig:o3_flux_AB_profile}). 
If one of the line-emitting regions were primarily illuminated by the distant AGN, the [O~{\sc iii}] emission would be expected to be dominated by the primary AGN and to decline broadly with distance from it, although this simple expectation may be affected by gas-density variations, dust attenuation, and projection effects.
In ESO 509-066, Mrk 463, Mrk 739, J0051+0020, J1414$-$0000, J1544+0446, and NGC 1128, the [O~{\sc iii}] profiles show local enhancements at the positions of both nuclei.  
These results suggest that the ionized gas around each nucleus is likely powered, at least in part, by a local ionizing source.
In IC 1623, the nucleus A show a [O~{\sc iii}] deficit, which may be caused by dust obscuration. 
Previous studies have suggested that IC 1623A hosts a dust-enshrouded starburst and a heavily obscured AGN \citep{2001AJ....122.1213S,2022ApJ...940L...8E,2023ApJ...944L..50R,2024ApJ...966..166B}.
For IC 1623B, J0853+1626B, NGC 3341A, and NGC 7592A, no significant [O~{\sc iii}] peaks are seen in their surface brightness profiles. 
These four nuclei are classified as composite regions in the BPT diagram, which may indicate contamination from circumnuclear SF.

Based on our spatially resolved BPT diagnostics, both two nuclei in 5 of the 12 systems (ESO 509-066, Mrk 463, J0858+1822, J1544+0446, and NGC 1128) are classified as Seyfert- or LI(N)ER ionization (Table~\ref{tab:bpt table}).
We note that J0858+1822 shows only a single [O~{\sc iii}] peak located between the two nuclei in Figure~\ref{fig:o3_flux_AB_profile}, consistent with the result of \citet{2020A&A...639A.117H}. 
This morphology suggests that the ionization structure in J0858+1822 may have a more complex origin. 
We therefore exclude this system from the following discussion.
Specifically, ESO 509-066 and Mrk463 are Seyfert+Seyfert pairs, whereas J1544+0446 and NGC 1128 are Seyfert+LI(N)ER or LI(N)ER+LI(N)ER systems, respectively. 
These four systems also exhibit local [O~{\sc iii}] enhancements at the positions of both nuclei.
Previous X-ray and/or radio observations have independently identified ESO 509-066 \citep{2005A&A...429L...9G,2017ApJ...850..168K}, Mrk 463 \citep{2008MNRAS.386..105B}, J1544+0446 \citep{2019ApJ...882...41H}, and NGC 1128 \citep{1985ApJ...294L..85O,2006A&A...453..433H} as dual AGNs.
Although LI(N)ER emission can also be powered by non-AGN processes, such as evolved stellar populations or shock excitation \citep[e.g.,][]{2010MNRAS.403.1036C,2014MNRAS.444.3894H}, these multi-wavelength results support the interpretation that these four systems host genuine AGN pairs.

Optically, the remaining 8 systems appear as single AGNs based on the BPT maps. 
Among these systems, Mrk 739, J0051+0020, and J1414$-$0000 all show local [O~{\sc iii}] enhancements at the positions of both nuclei. 
In addition, previous X-ray and/or radio observations suggest that these three systems may host genuine AGN pairs, including Mrk 739 \citep{2011ApJ...735L..42K}, J0051+0020 \citep{2015ApJ...799...72F,2015ApJ...815L...6F,2019ApJ...883...50G}, and J1414$-$0000 \citep{2020ApJ...900...79H}. 
The difference between the optical BPT classifications and the X-ray/radio results highlights the well-known limitations of optical emission-line diagnostics in interacting systems \citep[e.g.,][]{2008ApJ...677..926S,2009MNRAS.398.1165G,2013ApJ...764..176J,2014MNRAS.441.1297S,2017MNRAS.468.1273R}.
Optical AGN signatures can be masked by several factors: the dilution of AGN line ratios by intense circumnuclear star formation \citep[e.g.,][]{2015ApJ...811...26T}, heavy dust obscuration suppressing high-ionization lines \citep[e.g.,][]{2009MNRAS.398.1165G}, or an intrinsically low AGN luminosity relative to the host galaxy \citep[e.g.,][]{2008ARA&A..46..475H, 2017ApJ...835..223S}. 
In such scenarios, an AGN may dominate at high energies while remaining optically hidden, underscoring the crucial role of multi-wavelength approaches in accurately conducting a census of dual AGNs in galaxy mergers.

\subsection{Kinematics of the extended ionized gas} \label{subsec:extended gas}

The spatially resolved kinematic maps of our 12 targets, derived from MUSE data, typically exhibit complex features. 
These features indicate the combined effects of gravitational interactions and feedback processes. 
Table~\ref{tab:gas_structure} summarizes the gas structures of the extended ionized gas observed in our sample.

We detect regularly rotating ionized gas disks around 16 nuclei (Table~\ref{tab:gas_structure}). 
These disks are primarily traced by the first gas component, which is characterized by the rotational velocity field and relatively low velocity dispersion. 
In 6 of the 12 pairs (ESO 509-066, IC1623, J0051+0020, J0853+1626, J1414$-$0000, and NGC 7592), we observe distinct gas disks around both nuclei. 
In Mrk 739, J0858+1822, and NGC 3341, only Mrk 739A, J0858+1822A, and NGC 3341A present ionized gas disks.
The high fraction of nuclei retaining regularly rotating gas disks (16/24, $67\%$) is broadly consistent with IFU surveys of interacting galaxies at comparable merger stages. 
\cite{2013A&A...557A..59B} and \cite{2015ApJ...803...62H} found that $\sim40$–-$70\%$ of local (U)LIRGs in the interaction phase still exhibit rotation-dominated kinematics.
This result is also consistent with hydrodynamical simulations of galaxy mergers \citep[e.g.][]{1996ApJ...471..115B,1996ApJ...464..641M,2006ApJS..163....1H}. 
These simulations show that gas disks can remain largely unperturbed during the initial approach and may partially preserve or recover ordered rotation after the first pericentric passage. 
At the same time, the outer disk can be stripped into tidal tails, while gravitational torques drive gas inflows toward the nuclear regions. 
At later stages, especially near final coalescence, simulations predict that the residual gas disk becomes increasingly turbulent and loses much of its ordered rotational support.

Clear signatures of tidal tails are also present. 
We observe prominent ionized gas tidal tails in 9 of the 12 pairs (ESO 509-066, IC 1623, Mrk 463, Mrk 739, J0853+1626, J0858+1822, J1544+0446, NGC 3341, and NGC 7592). 
The kinematic continuity between the rotating disks and the tidal tails suggests that the gas in these structures originated in the galactic disks and was gravitationally stripped during the interaction. 
This picture agrees well with hydrodynamical simulations, in which tidal tails form primarily from loosely bound outer-disk material removed during close passages, especially after first pericentric passage \citep[e.g.][]{1996ApJ...471..115B,1996ApJ...464..641M}.

In contrast to rotating disks and tidal features, 18 of the 24 nuclei exhibit signatures of ionized gas outflows. 
These outflows are identified from the presence of a second emission-line component, which is characterized by significantly higher $\sigma_{\rm 2nd}$ values and kinematic structures that deviate from simple disk rotation (see Sect~\ref{subsec:ESO 509-066}).
In J1414$-$0000 and NGC 1128, the outflowing gas is not spatially resolved. 
In IC 1623, Mrk 463, and J0858+1822, however, we observe prominent large-scale outflows that extend well beyond the nuclear regions and reach projected radial distances of more than 10 kpc. 
The high incidence of outflows in our sample is broadly consistent with merger simulations in which tidal torques drive gas inflows toward the central regions and enhance both nuclear SF and BH accretion \citep{1996ApJ...471..115B,2006ApJS..163....1H,2015MNRAS.447.2123C}.

\subsection{Evidence of fading AGNs} \label{subsec:fading AGN}

\begin{figure*}[ht!]
\includegraphics[width=0.95\textwidth,trim=0 0 0 0]{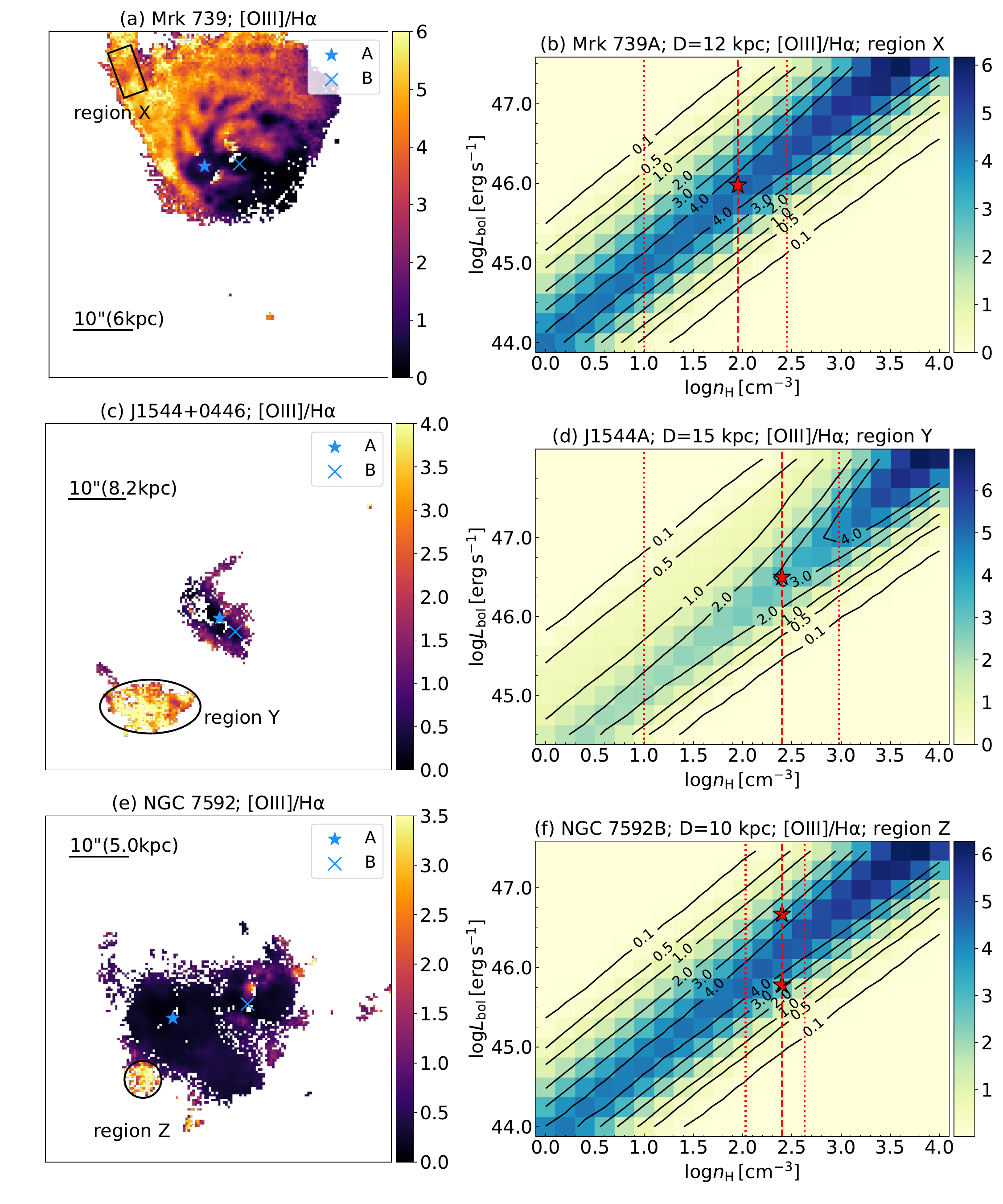}
\caption{(a): Map of the flux ratio [O~{\sc{iii}}]/H$\alpha$ for Mrk 739.
(b) Predicted [O~{\sc iii}]/H$\alpha$ ratios from CLOUDY photoionization models, assuming a typical AGN SED with an Eddington ratio of $L_{\rm AGN}/L_{\rm Edd}=0.76$ \citep[][]{2012MNRAS.420.1825J} and a cloud distance of 12 kpc from Mrk 739A.
The red stars mark the solutions defined by the intersection of the measured electron density (red dashed line) and the [O~{\sc iii}]/H$\alpha$ ratio derived for region X. 
The dotted lines indicate the $1\sigma$ uncertainty range of the electron density.
(c) and (d) are similar to (a) and (b), but for J1544+0446A and region Y.
(e) and (f) are similar to (a) and (b), but for NGC 7592B and region Z.
Detailed parameters are listed in Table~\ref{tab:luminosity_CLOUDY_xray}.}
\label{fig:cloudy_ratio}
\end{figure*}

\begin{table*}[]
\centering
\begin{tabular}{cccccccccc}
\hline
name & region & $D_{\rm region}$ & $F_{[\rm O~III]}/F_{\rm H\alpha}$ & $n_{\rm e}$ & $Z_{\rm region}/Z_{\odot}$ & $E(\rm B-V)$ & ${\rm log}(L_{\rm bol}^{\rm region})$ & ${\rm log}(L_{\rm X})$ & ${\rm log}(L_{\rm bol}^{\rm X})$ \\ 
(1) & (2) & (3) & (4) & (5) & (6) & (7) & (8) & (9) & (10) \\
\hline\hline
Mrk 739A    & X & 12 kpc & 5.1 & $90^{+190}_{-80}$ & 1.1 & 0.27 & $45.8^{+0.5}_{-0.8}$ & 43.04 & $44.11$ \\ \hline
J1544+0446A & Y & 15 kpc & 3.5 & $250^{+705}_{-240}$ & 2.1 & 0.10 & $46.5^{+0.4}_{-0.2}$ & 41.55 & $42.59$ \\ \hline
NGC 7592B   & Z & 10 kpc & 3.1 & $250^{+176}_{-142}$ & 1.0 & 0.32 & $45.8^{+0.2}_{-0.3}$ & 40.77 & $41.81$ \\ \hline
\end{tabular}
\caption{Detailed parameters associated with Figure~\ref{fig:cloudy_ratio}.
The columns are as follows: (1) Name of the nucleus. 
(2) Region name. 
(3) Projected distance of the region from the nucleus.
(4) [O~{\sc{iii}}]/H$\alpha$ flux ratio.
(5) Electron density ($\rm\,cm^{-3}$), derived from the [S~{\sc{ii}}]$\lambda\lambda 6716,6731$ doublet ratio measured in the region.
(6) Gas-phase metallicity derived from the [O~{\sc iii}]$\lambda\lambda4959,5007$/H$\beta$ and [N~{\sc ii}]$\lambda\lambda6548,6584$/H$\alpha$ using the method of \citep[][]{1998AJ....115..909S}.
(7) $E(\rm B-V)$ estimated from the Balmer decrement ($\rm H\alpha/H\beta$.)}
(8) Bolometric luminosity ($\rm\,erg\,s^{-1}$) required to photoionize the gas cloud in the region, as predicted by CLOUDY.
(9) Hard X-ray (2–-10 keV) luminosity ($\rm\,erg\,s^{-1}$) of Mrk 739A, J1544+0446A, and NGC 7592B, taken from \cite{2011ApJ...735L..42K}, \cite{2019ApJ...882...41H}, and \cite{2018A&A...620A.140T}, respectively.
(10) Bolometric luminosity ($\rm\,erg\,s^{-1}$) converted from the hard X-ray luminosity using the bolometric correction of \cite{2020A&A...636A..73D}, with an associated uncertainty of $\sim0.27$ dex.
\label{tab:luminosity_CLOUDY_xray}
\end{table*}

Observations show that some AGNs can fade by orders of magnitude within $\sim 1$--$10\times10^{4}\rm\, yr$ \citep[e.g.][]{2012MNRAS.420..878K,2017ApJ...835..256K,2022ApJ...938...75P,2025ApJ...993...35X}.
The galaxies in our sample are all merging systems, in which the central AGN activity may undergo more rapid and violent variations. 
Very extended ionized gas features ($>10\rm\, kpc$) are found in some of these galaxies, and portions of these structures are classified as AGN-photoionized on the BPT diagrams (e.g., Mrk 739, J1544+0446, and NGC 7592), implying the presence of luminous AGNs capable of ionizing the gas at such large distances.

In Mrk 739, NGC 7592, and J1544+0446, we find evidence for AGN fading over the past several $10^{4}\rm\, yr$, following a method similar to that of \cite{2025ApJ...993...35X}. 
Using CLOUDY \citep[v23.01,][]{2023RMxAA..59..327C,2023RNAAS...7..246G} with an AGN SED characterized by a high Eddington ratio ($L_{\rm AGN}/L_{\rm Edd}=0.76$) from \cite{2012MNRAS.420.1825J}, we reproduced the observed [O~{\sc iii}]/H$\alpha$ ratios in three selected regions (regions X, Y, and Z) across these three galaxies (Figure~\ref{fig:cloudy_ratio}).
According to the BPT diagrams (Figure~\ref{fig:Mrk739_flux_kinematics_bpt}, Figure~\ref{fig:N7592_flux_kinematics_bpt}, and Figure~\ref{fig:J1544_flux_kinematics_bpt}), these three regions are dominated by the AGN photoionization, and local SF cannot account for their high line ratios. 
Moreover, because the velocity dispersions in these regions are low ($\sigma_{\rm 1st}<70\rm\,km\,s^{-1}$), shocks are unlikely to be the primary excitation mechanism.
We estimate the dust extinction, $E(\rm B-V)$, in these three regions using the Balmer decrement. 
All emission lines are corrected for dust extinction by adopting the attenuation law of \citet{2000ApJ...533..682C} with $\rm R_{V}=3.1$.
The derived $E(\rm B-V)$ values are low in all three selected regions (Table~\ref{tab:luminosity_CLOUDY_xray}), indicating that dust extinction is not significant.
The electron densities are estimated from the [S~{\sc ii}]$\lambda\lambda 6716,6731$ doublet ratio of the first gas component.

We estimate the gas-phase metallicities of the selected regions, namely regions X, Y, and Z, using the calibration of \citet{1998AJ....115..909S}, which is based on photoionization models for AGNs. 
This method uses the optical emission-line ratios [O~{\sc iii}]$\lambda\lambda4959,5007$/H$\beta$ and [N~{\sc ii}]$\lambda\lambda6548,6584$/H$\alpha$ to infer the $12+{\rm log(O/H)}$. 
The calibration is appropriate for AGN-like narrow-line regions and is valid over the metallicity range $8.4\leq12+{\rm log(O/H)}\leq9.2$.
We adopt the solar metallicity of $12+{\rm log(O/H)}\approx8.69$ from \citep[][]{2009ARA&A..47..481A}.
The metallicities of the three selected regions, expressed as$Z_{\rm region}/Z_{\odot}$, are summarized in Table~\ref{tab:luminosity_CLOUDY_xray}.

Region X in Mrk 739 (Figure~\ref{fig:cloudy_ratio}) is dominated by AGN photoionization and exhibits a high [O~{\sc iii}]/H$\alpha$ ratio ($F_{[\rm O~III]}/F_{\rm H\alpha}=5.1$), indicating that a luminous AGN is responsible for ionizing the gas.
Based on the hard X-ray luminosity of Mrk 739A, which is higher than that of Mrk 739B, we assume that the ionizing photons illuminating region X originate primarily from Mrk 739A.
As shown in Figure~\ref{fig:cloudy_ratio}, reproducing the ionization state of region X requires an AGN bolometric luminosity of ${\rm log}(L_{\rm bol}^{\rm region\,X}/{\rm erg\,s^{-1}}) = 45.8^{+0.5}_{-0.8}$ for Mrk 739A, given an electron density of $n_{\rm e}=90^{+190}_{-80}\rm\, cm^{-3}$.
The lower and upper limits of ${\rm log}(L_{\rm bol}^{\rm region\,X})$ are determined from the intersections of the corresponding lower and upper limits of the electron density and the $F_{[\rm O~III]}/F_{\rm H\alpha}$ ratio.
The [S~{\sc ii}]$\lambda\lambda 6716,6731$ doublet ratio is sensitive to electron densities over the approximate range $50\leq n_{\rm e} \leq5000\rm\, cm^{-3}$ \citep{2006agna.book.....O}. 
Since the estimated $n_{\rm e}$ for region X is relatively low, we adopt a lower limit of $n_{\rm e}=10\rm\, cm^{-3}$.
Given that the metallicity of region X is $Z_{\rm region}/Z_{\odot}\sim1.1$, we adopt the solar metallicity from \citet{2009ARA&A..47..481A} in our CLOUDY modeling.
From the hard X-ray luminosity of Mrk 739A \citep[${\rm log}(L_{\rm X}/{\rm erg\,s^{-1}})=43.04$,][]{2011ApJ...735L..42K}, the current bolometric luminosity is estimated to be ${\rm log}(L_{\rm bol}^{\rm X}/{\rm erg\,s^{-1}})=44.11^{+0.27}_{-0.27}$, adopting the bolometric correction with an associated uncertainty of $\sim0.27\rm\,dex$ from \cite{2020A&A...636A..73D}.
These parameters are listed in Table~\ref{tab:luminosity_CLOUDY_xray}.
The inferred $L_{\rm bol}^{\rm region\,X}$ is approximately 30 times larger than $L_{\rm bol}^{\rm X}$ for Mrk 739A. 
The projected distance from Mrk 739A to region X is $\sim12\rm\,kpc$ corresponding to a light travel time of $\sim40,000\rm\, yr$.
This implies that Mrk 739A was more luminous $\sim40,000\rm\, yr$ ago than it is at present.
For $n_{\rm e}=10$--$100\,{\rm cm^{-3}}$, the recombination timescale of ${\rm O}^{++}$ is sufficiently short compared with the relevant light-travel time. 
We therefore neglect recombination delays and consider only light-travel-time effects.

Similarly, the required AGN luminosity can be estimated for region Y in J1544+0446. 
Based on Figure~\ref{fig:cloudy_ratio}(d), the AGN luminosity is estimated to be ${\rm log}(L_{\rm bol}^{\rm region\,Y}/{\rm erg\,s^{-1}})\sim46.5^{+0.4}_{-0.2}$.
Although the lower limit on the electron density of region Y is low, the observed high $F_{[\rm O~III]}/F_{\rm H\alpha}$ ratio requires ${\rm log}(L_{\rm bol}^{\rm region\,Y})\gtrsim46.3$.
A lower ${\rm log}(L_{\rm bol}^{\rm region\,Y})$ cannot reproduce the observed $F_{[\rm O~III]}/F_{\rm H\alpha}$ ratio in region Y.
Given that the metallicity of region Y is $Z_{\rm region}/Z_{\odot}\sim2.1$, we adopt twice the solar metallicity from \citet{2009ARA&A..47..481A} in our CLOUDY modeling.
Given the hard X-ray luminosity of J1544+0446A \citep[${\rm log}(L_{\rm X}/{\rm erg\,s^{-1}})=41.55$,][]{2019ApJ...882...41H}, the current AGN bolometric luminosity of J1544+0446A is ${\rm log}(L_{\rm bol}^{\rm X}/{\rm erg\,s^{-1}})=42.59^{+0.27}_{-0.27}$.
This result indicates that the AGN in J1544+0446A must have been brighter by more than 3--4 orders of magnitude $\sim50,000\rm\, yr$ ago.

For region Z in NGC 7592, adopting a projected distance of $D=10\rm\,kpc$ from NGC 7592B to region Z, the required AGN luminosity is ${\rm log}(L_{\rm bol}^{\rm region\,Z}/{\rm erg\,s^{-1}})\sim45.8^{+0.2}_{-0.3}$, while the current AGN bolometric luminosity of NGC 7592B is ${\rm log}(L_{\rm bol}^{\rm X}/{\rm erg\,s^{-1}})=41.81^{+0.27}_{-0.27}$.
Given that the metallicity of region Y is $Z_{\rm region}/Z_{\odot}\sim1.0$, we adopt the solar metallicity from \citet{2009ARA&A..47..481A} in our CLOUDY modeling.
Although NGC 7592A is physically closer to region Z, the BPT diagram (Figure~\ref{fig:N7592_flux_kinematics_bpt}) shows that NGC 7592A is dominated by SF, whereas NGC 7592B exhibits AGN photoionization.
This result suggests that the AGN in NGC 7592B must have been more luminous by more than 2--3 orders of magnitude $\sim40,000\rm\, yr$ ago.

In the above, we adopt an AGN SED with a high Eddington ratio ($L_{\rm AGN}/L_{\rm Edd}=0.76$) from \cite{2012MNRAS.420.1825J} as the input SED for estimating the required AGN luminosities. 
If an AGN SED with a lower Eddington ratio were adopted, the derived AGN luminosity would be somewhat higher than our current estimate of $L_{\rm bol}^{\rm region}$.
Consequently, the use of a high-Eddington-ratio AGN SED may lead to a slight underestimate of the true AGN luminosity, but this does not affect the validity of our conclusions.

Extended ionized gas features are also found in ESO 509-066 and Mrk 463. 
We estimate the AGN bolometric luminosities in these galaxies using the same CLOUDY modeling approach, and find no significant differences from their current AGN luminosities inferred from hard X-ray observations, with $\log(L_{\rm bol}^{\rm X}/{\rm erg\,s^{-1}})\sim44.6$ for ESO 509-066A and $\log(L_{\rm bol}^{\rm X}/{\rm erg\,s^{-1}})\sim44.3$ for Mrk 463A. 
These results suggest that the AGNs in these sources could have maintained a relatively stable luminosity over the past $\sim30,000$--$50,000\rm\, yr$.

We also consider whether the ionizing-luminosity deficit could be caused by directional nuclear emission. 
Relativistic beaming could in principle make the ionizing radiation stronger toward the extended clouds than along our line of sight. 
However, such beaming is usually associated with a radio jet or a collimated relativistic outflow.
Compact nuclear radio emission has been detected in Mrk 739A \citep[][]{2011ApJ...735L..42K}, while no radio observations have been reported for J1544+0446 and NGC 7592.
We cannot completely exclude weak or unresolved jet activities in these three galaxies, but in the absence of any observed jet-like morphology or alignment with the extended clouds, relativistic beaming would require a finely tuned geometry and is unlikely to dominate the ionizing radiation received by the clouds.

A more plausible possibility is anisotropic illumination, in which ionizing photons escape preferentially along certain directions, for example through an ionization cone. 
In this case, the clouds could receive a larger ionizing flux than inferred from the current X-ray luminosity. 
The required anisotropy factor is approximately $A_{\rm ion}=L_{\rm bol}^{\rm region}/L_{\rm bol}^{\rm X}$.
In a simple biconical illumination model, the required anisotropy factor is $A_{\rm ion}\sim4\pi/\Omega_{\rm cone}$.
For a bicone with half-opening angle $\theta$, this gives $A_{\rm ion}\sim(1-\rm cos\,\theta)^{-1}$.
Thus, an anisotropy factor of $A_{\rm ion}\geq30$ would require $\theta\leq15^{\circ}$, corresponding to a full opening angle of $\leq30^{\circ}$.
Larger luminosity deficits would require even narrower cones.
For Mrk 739A, we find a projected full opening angle of $\sim30^{\circ}$ for region X, which is consistent with the angle required for an anisotropy factor of $A_{\rm ion}\sim30$.
This result suggests that anisotropic illumination could explain the ionizing-luminosity deficit in Mrk 739A.
However, for NGC 7592B and J1544+0446A, the required anisotropy factors are $A_{\rm ion}>100$, corresponding to full opening angles of $<16^{\circ}$.
These required opening angles are much smaller than the observed values for the two sources. 
Therefore, anisotropic illumination alone cannot fully account for the ionizing-luminosity deficits in NGC 7592B and J1544+0446A. 

\section{Summary} \label{sec:summary}

We have presented a spatially resolved study of the ionized gas kinematics and ionization properties for 12 AGN pair candidates in the local universe ($z\lesssim0.1$), utilizing archival data from VLT/MUSE. 
By decomposing the optical emission lines into two Gaussian components, we try to disentangle the gas associated with disk rotation and tidal features from non-circular motions.
Furthermore, we employed spatially resolved BPT diagrams to identify the dominant ionization mechanisms across the systems.
The main results and conclusions are summarized as follows:

\begin{enumerate}

    \item Using spatially resolved BPT maps, we find that in 4 of the 12 systems (ESO 509-066, Mrk 463, NGC 1128, and J1544+0446), both nuclei are dominated by AGN-like ionization, being classified as either Seyfert or LI(N)ER (see Table~\ref{tab:bpt table}). 
    Three nuclei (Mrk 739B, J0051+0020A, and J1414$-$0000A) are classified as SF or composite in the optical BPT diagram, but have been identified as AGNs at other wavelengths (Section~\ref{subsec:nature of AGN-pair}), suggesting that these three systems are also AGN pairs.
    
    \item The kinematic analysis shows that regular rotation is the predominant feature in our sample. 
    In 16 of the 24 nuclei, we detect a regularly rotating ionized gas disk, primarily traced by the narrower emission-line component (the first component), which is characterized by low $\sigma_{\rm 1st}$ and rotational velocity field (Table~\ref{tab:gas_structure}). 
    Moreover, 6 of the 12 pairs (ESO 509-066, IC 1623, J0051+0020, J0853+1626, J1414$-$0000, and NGC 7592) host dual gas disks associated with their respective nuclei. 
    Prominent tidal features are also observed in 9 targets (ESO 509-066, IC 1623, Mrk 463, Mrk 739, J0853+1626, J0858+1822, J1544+0446, NGC 3341, and NGC 7592). 
    In addition, ionized gas outflows are widespread, being detected in 18 of the 24 nuclei and primarily traced by the broader emission-line component (the second component).
    
    \item Based on the optical emission line ratios and the CLOUDY modeling (Sect.~\ref{subsec:fading AGN}), we find evidence for fading AGN activity over the past several $10^{4}\rm\, yr$ in Mrk 739, J1544+0446, and NGC 7592. 
    In Mrk 739, the AGN in Mrk 739A is inferred to have been more luminous $\sim40,000\rm\, yr$ ago, with ${\rm log}(L_{\rm bol,region\,A}) = 45.8^{+0.5}_{-0.8}\rm\,erg\,s^{-1}$, about one order of magnitude higher than its current bolometric luminosity.
    Alternatively, anisotropic illumination could account for the ionizing-luminosity deficit in Mrk 739A if the full opening angle is $\leq30^{\circ}$ (Sect~\ref{subsec:fading AGN}).
    In J1544+0446, the AGN in J1544+0446A is inferred to have been significantly brighter $\sim50,000\rm\, yr$ ago, with ${\rm log}(L_{\rm bol}^{\rm region\,Y})\sim46.2^{+0.6}_{-0.2}\rm\,erg\,s^{-1}$, exceeding the current luminosity by more than 3--4 orders of magnitude.
    Similarly, in NGC 7592, the AGN in NGC 7592B is inferred to have been more luminous $\sim40,000\rm\, yr$ ago, with ${\rm log}(L_{\rm bol}^{\rm region\,Z})\sim45.8^{+0.2}_{-0.3}\rm\,erg\,s^{-1}$, more than 2--3 orders of magnitude higher than its current bolometric luminosity.
    For J1544+0446A and NGC 7592B, anisotropic illumination cannot explain the ionizing-luminosity deficits, because the required full opening angles are too small.

\end{enumerate}


\begin{acknowledgments}

We thank the anonymous referee for helpful comments that significantly improved the clarity of our work. 
X.X. acknowledges the NSFC grant 12403018, the China Postdoctoral Science Foundation (No. 2023M741639), and the Jiangsu Funding Program for Excellent Postdoctoral Talent (No. 2024ZB249).
Z.L. acknowledges the National Natural Science Foundation of China (grant 12225302).
M.H. acknowledges support from the NSFC grant 12203001.
J.W. acknowledges the National Key R\&D Program of China (grant No. 2023YFA1607904), the NSFC grants 12333002, 12033004, 12221003, and China Manned Space Project with No. CMS-CSST-2025-A07 and CMS-CSST-2025-A10.
Supported by the Fundamental Research Funds for the Central Universities (KG202502).

Based on observations collected at the European Southern Observatory under ESO program 0103.A-0637,0100.B-0116, 110.23WR, 109.238W, 095.B-0482, and 0101.D-0748.

\end{acknowledgments}





%
\facilities{VLT:MUSE}

\software{astropy \citep{2013A&A...558A..33A,2018AJ....156..123A,2022ApJ...935..167A},  PPXF \citep{2004PASP..116..138C,2023MNRAS.526.3273C}, DS9 \citep{2003ASPC..295..489J}
          }


\appendix

\section{Surface brightness, kinematics, and BPT maps} \label{sec:maps_others}

Here we present the surface brightness, kinematic, and BPT maps for the remaining systems, with the exception of ESO509-066.

\begin{figure*}[ht!]
\includegraphics[width=1\textwidth,trim=0 0 0 0]{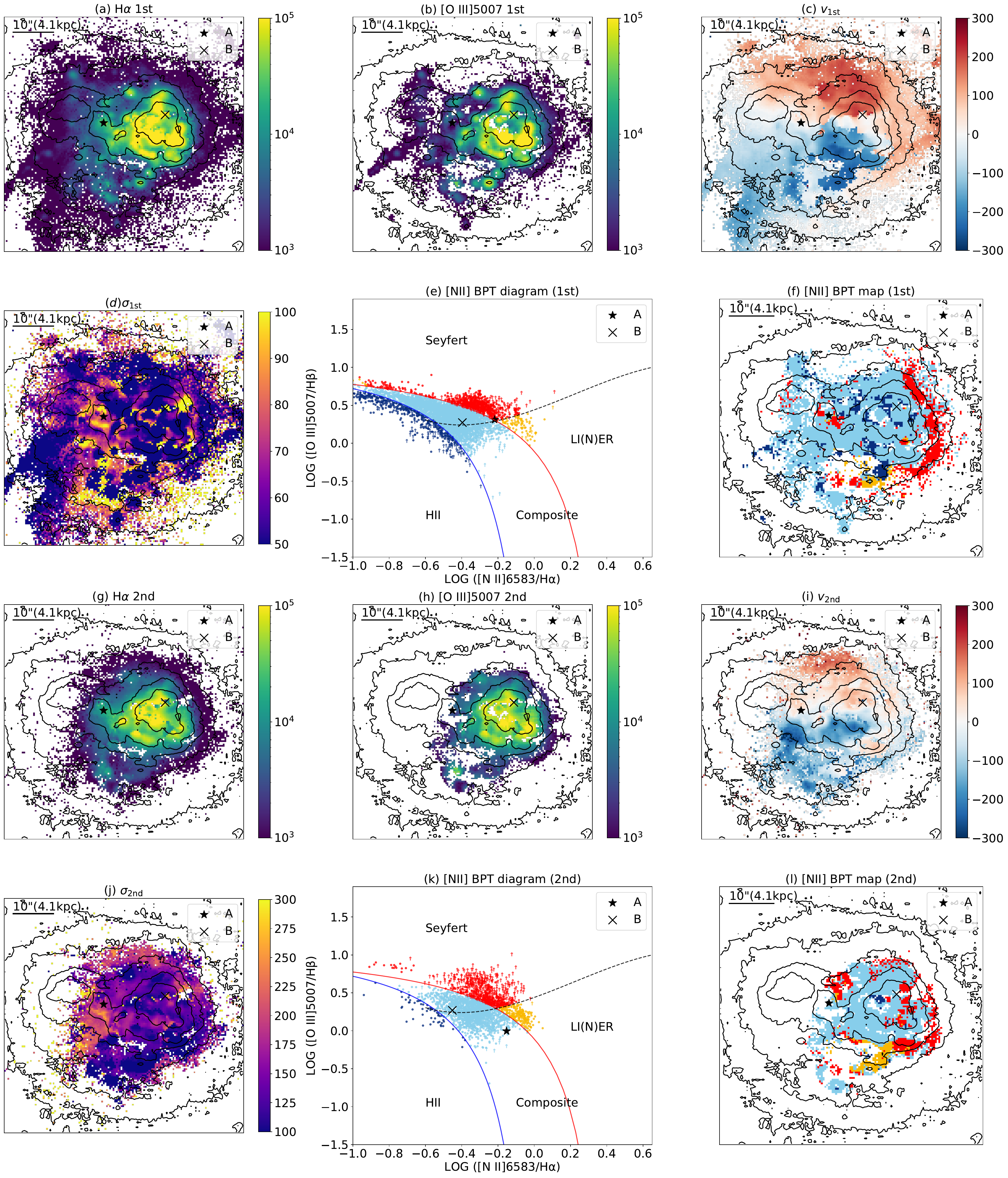}
\caption{Surface brightness, kinematics, and BPT maps of IC 1623. 
Similar to Figure~\ref{fig:E509_flux_kinematics_bpt}.
Velocities in $v_{\rm 1st}$ and $v_{\rm 2nd}$ maps are referenced to the first component of nucleus A.
}
\label{fig:IC1623_flux_kinematics_bpt}
\end{figure*}

\begin{figure*}[ht!]
\includegraphics[width=1\textwidth,trim=0 0 0 0]{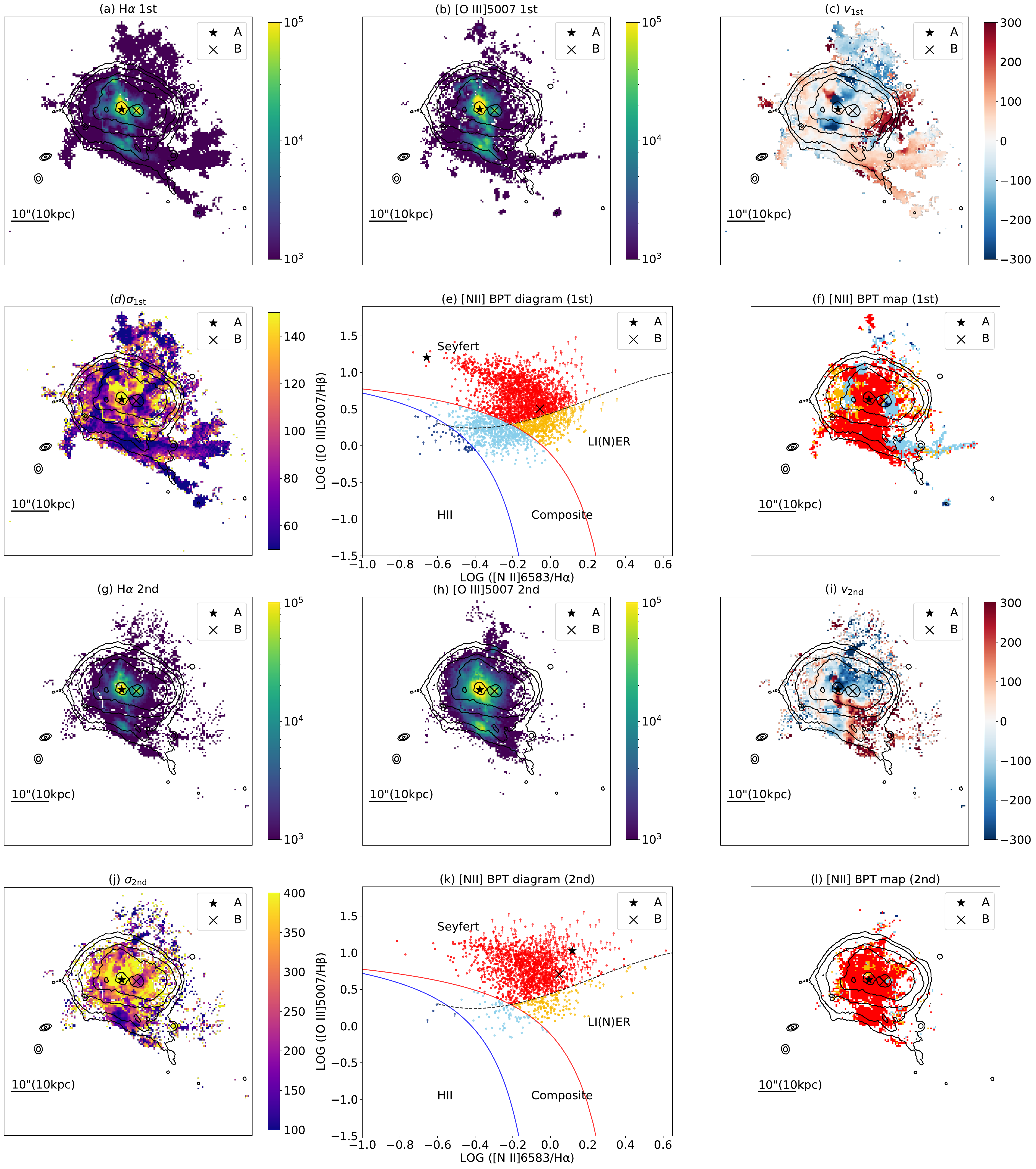}
\caption{Surface brightness, kinematics, and BPT maps of Mrk 463. 
Similar to Figure~\ref{fig:E509_flux_kinematics_bpt}.
Velocities in $v_{\rm 1st}$ and $v_{\rm 2nd}$ maps are referenced to the first component of nucleus A.
}
\label{fig:Mrk463_flux_kinematics_bpt}
\end{figure*}

\begin{figure*}[ht!]
\includegraphics[width=1\textwidth,trim=0 0 0 0]{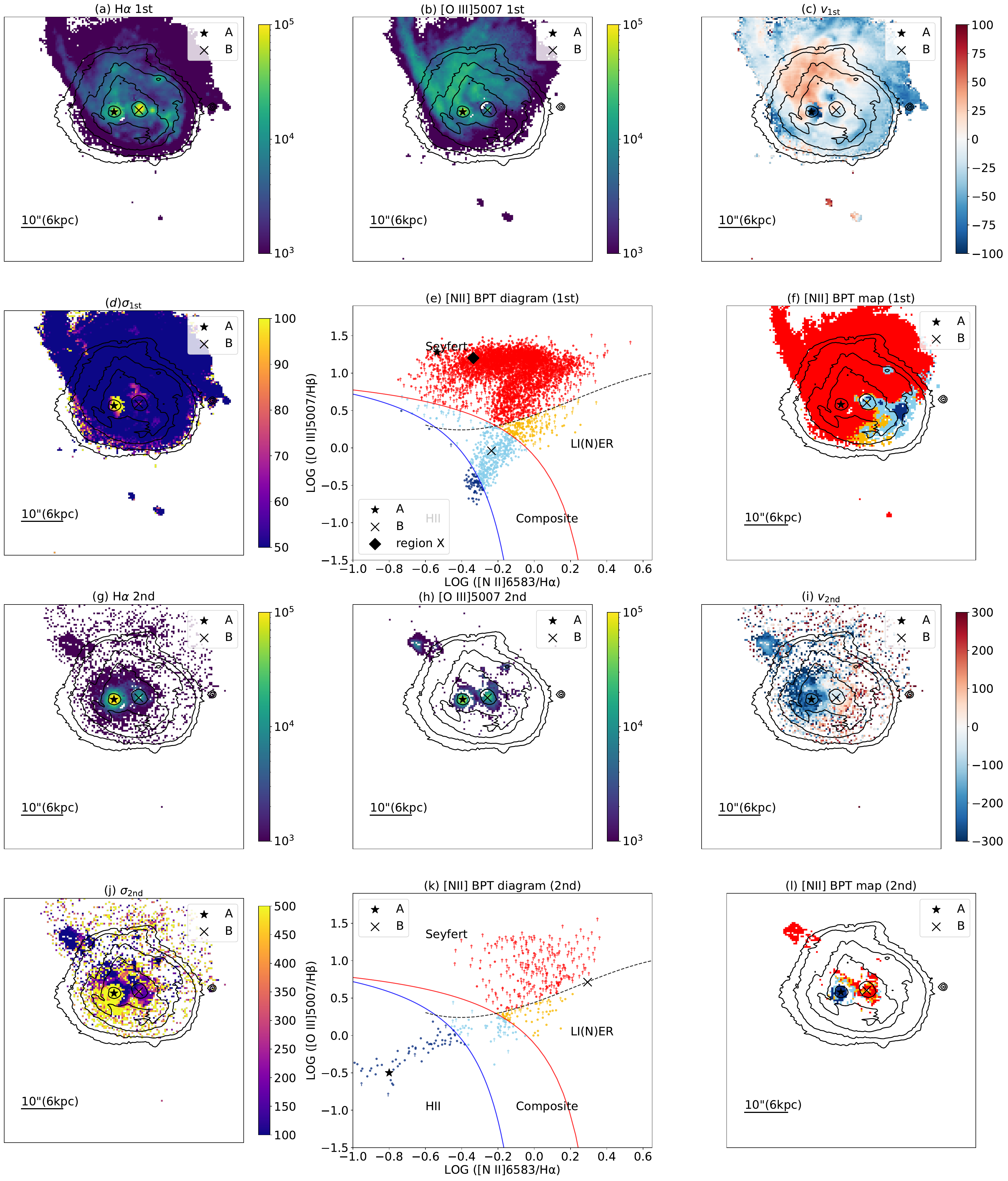}
\caption{Surface brightness, kinematics, and BPT maps of Mrk 739. 
Similar to Figure~\ref{fig:E509_flux_kinematics_bpt}.
Velocities in $v_{\rm 1st}$ and $v_{\rm 2nd}$ maps are referenced to the first component of nucleus B.
}
\label{fig:Mrk739_flux_kinematics_bpt}
\end{figure*}

\begin{figure*}[ht!]
\includegraphics[width=1\textwidth,trim=0 0 0 0]{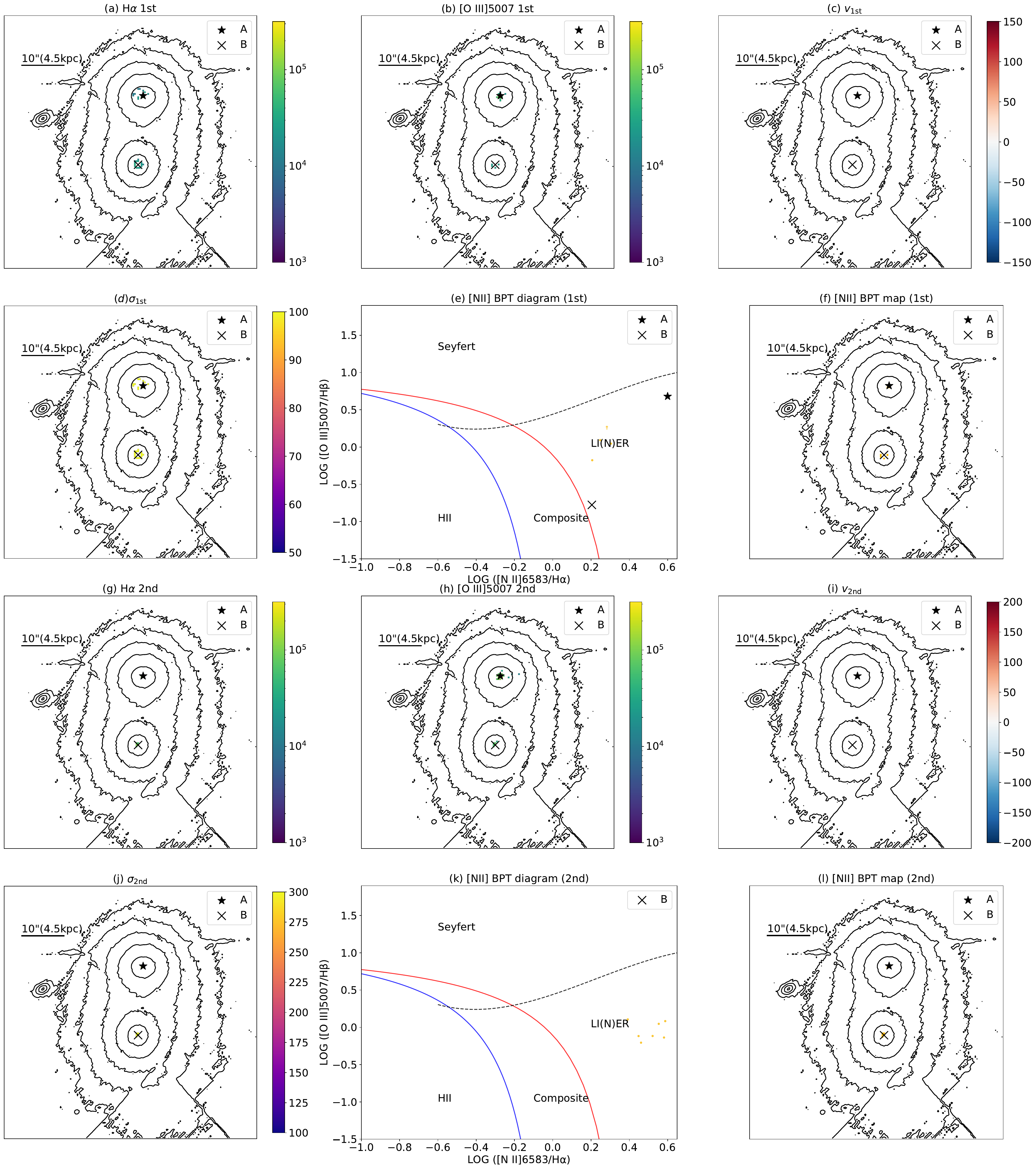}
\caption{Surface brightness, kinematics, and BPT maps of NGC 1128. 
Similar to Figure~\ref{fig:E509_flux_kinematics_bpt}.
Line emission is detected only in the two nuclear regions.
}
\label{fig:N1128_flux_kinematics_bpt}
\end{figure*}

\begin{figure*}[ht!]
\includegraphics[width=1\textwidth,trim=0 0 0 0]{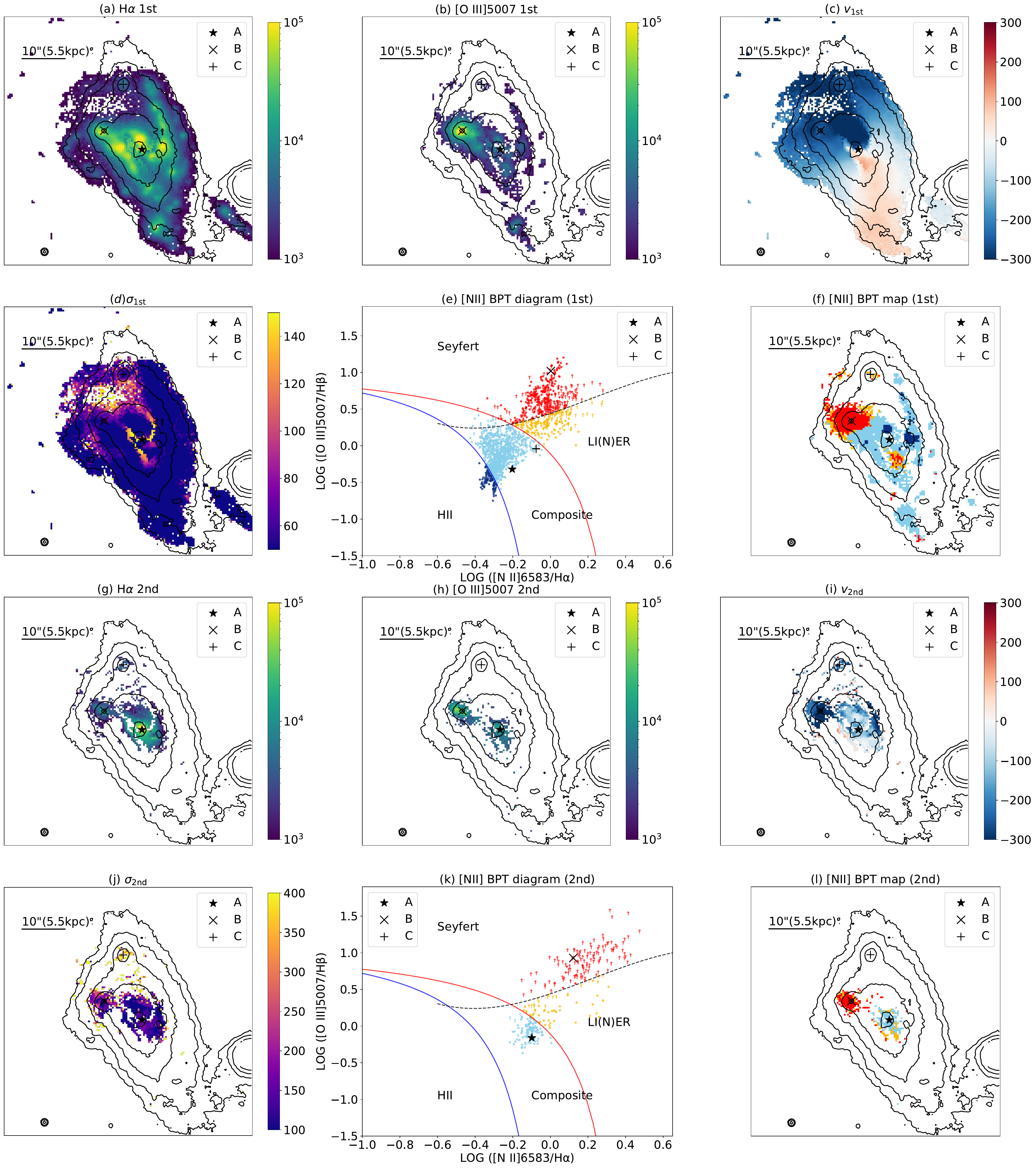}
\caption{Surface brightness, kinematics, and BPT maps of NGC 3341. 
Similar to Figure~\ref{fig:E509_flux_kinematics_bpt}.
Velocities in $v_{\rm 1st}$ and $v_{\rm 2nd}$ maps are referenced to the first component of nucleus A.
}
\label{fig:N3341_flux_kinematics_bpt}
\end{figure*}

\begin{figure*}[ht!]
\includegraphics[width=1\textwidth,trim=0 0 0 0]{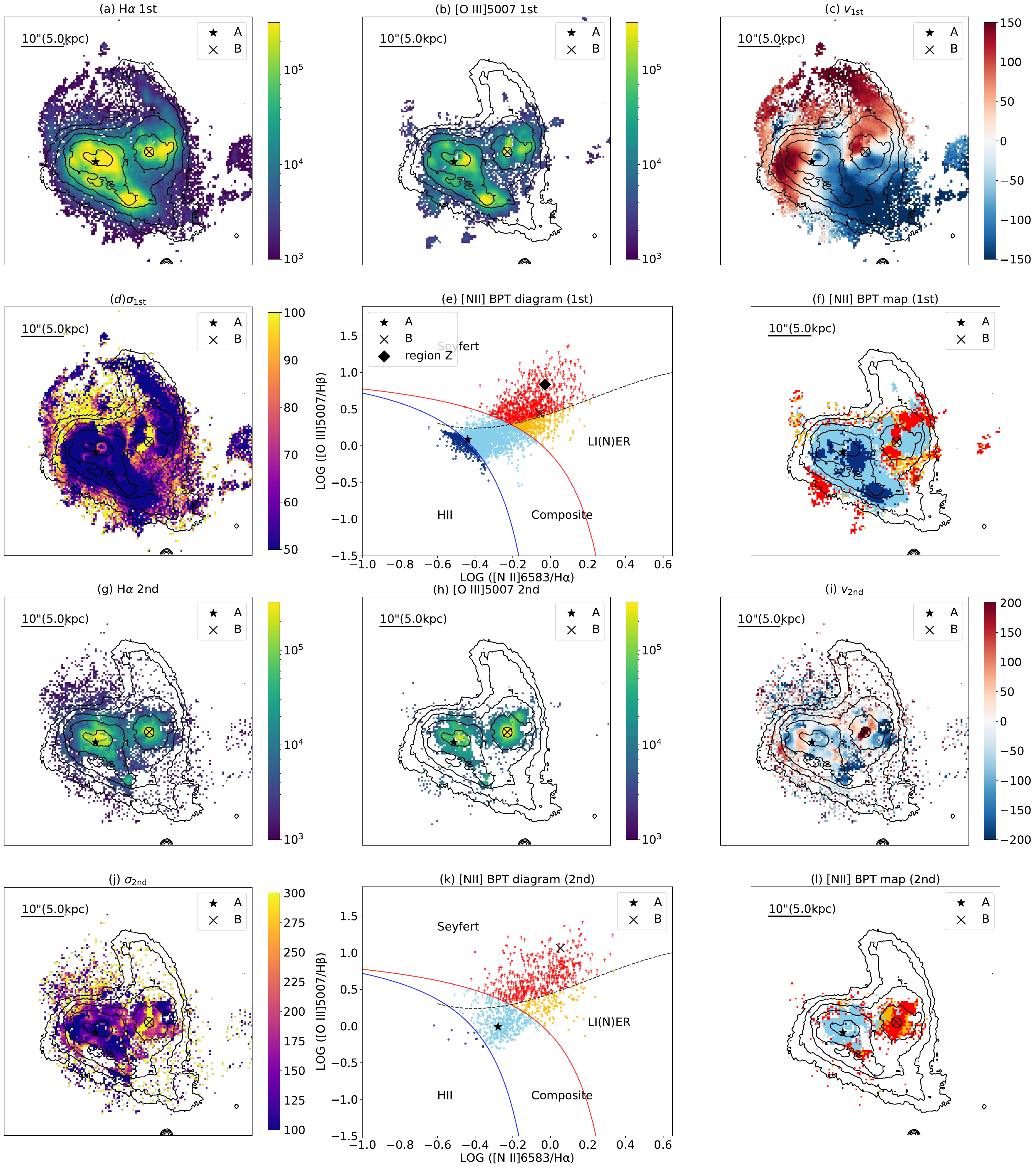}
\caption{Surface brightness and kinematics maps of NGC 7592. 
Similar to Figure~\ref{fig:E509_flux_kinematics_bpt}.
Velocities in $v_{\rm 1st}$ and $v_{\rm 2nd}$ maps are referenced to the first component of nucleus A.
}
\label{fig:N7592_flux_kinematics_bpt}
\end{figure*}

\begin{figure*}[ht!]
\includegraphics[width=1\textwidth,trim=0 0 0 0]{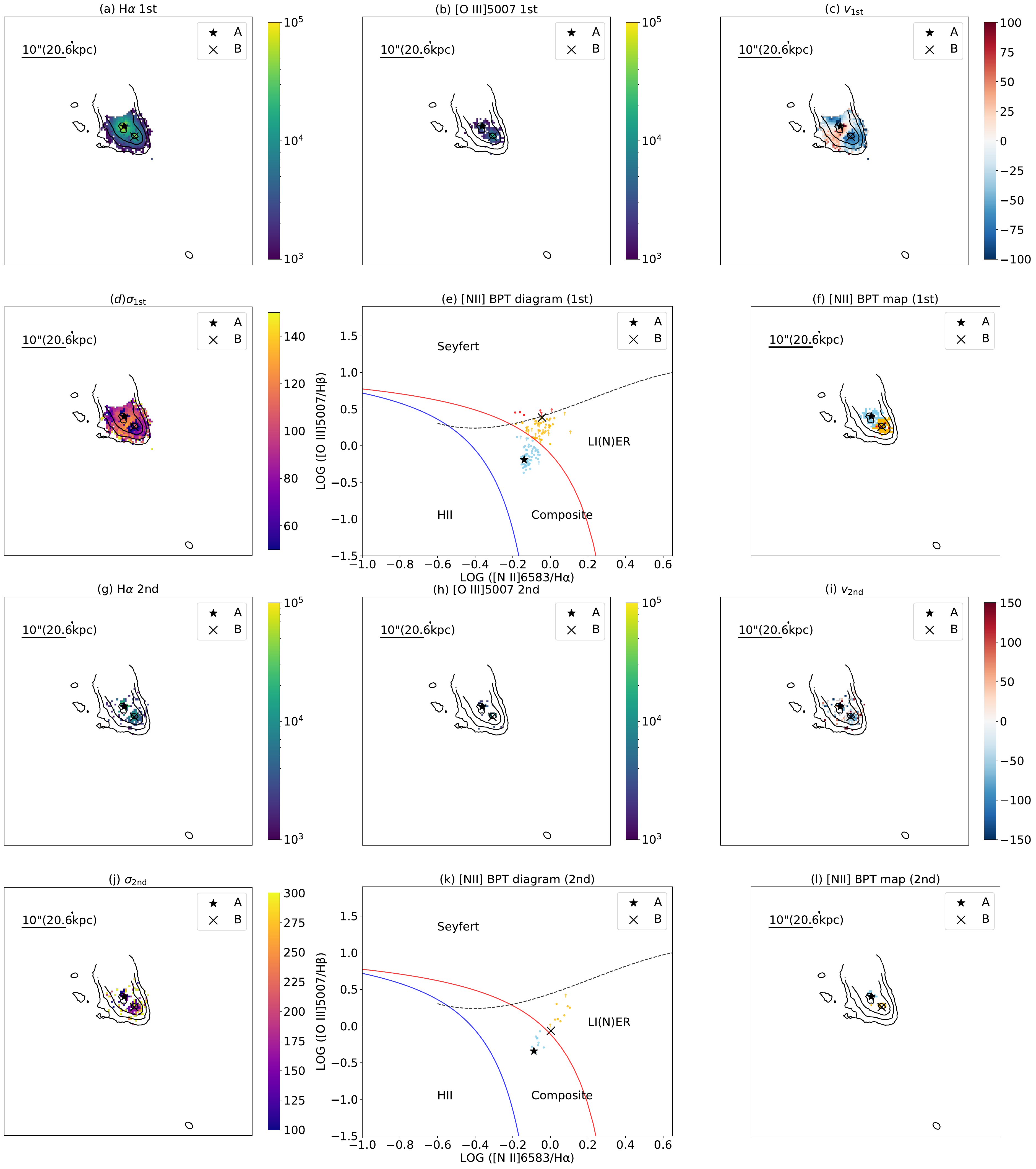}
\caption{Surface brightness, kinematics, and BPT maps of J0051+0020. 
Similar to Figure~\ref{fig:E509_flux_kinematics_bpt}.
Velocities in $v_{\rm 1st}$ and $v_{\rm 2nd}$ maps are referenced to the first component of nucleus B.
}
\label{fig:J0051_flux_kinematics_bpt}
\end{figure*}

\begin{figure*}[ht!]
\includegraphics[width=1\textwidth,trim=0 0 0 0]{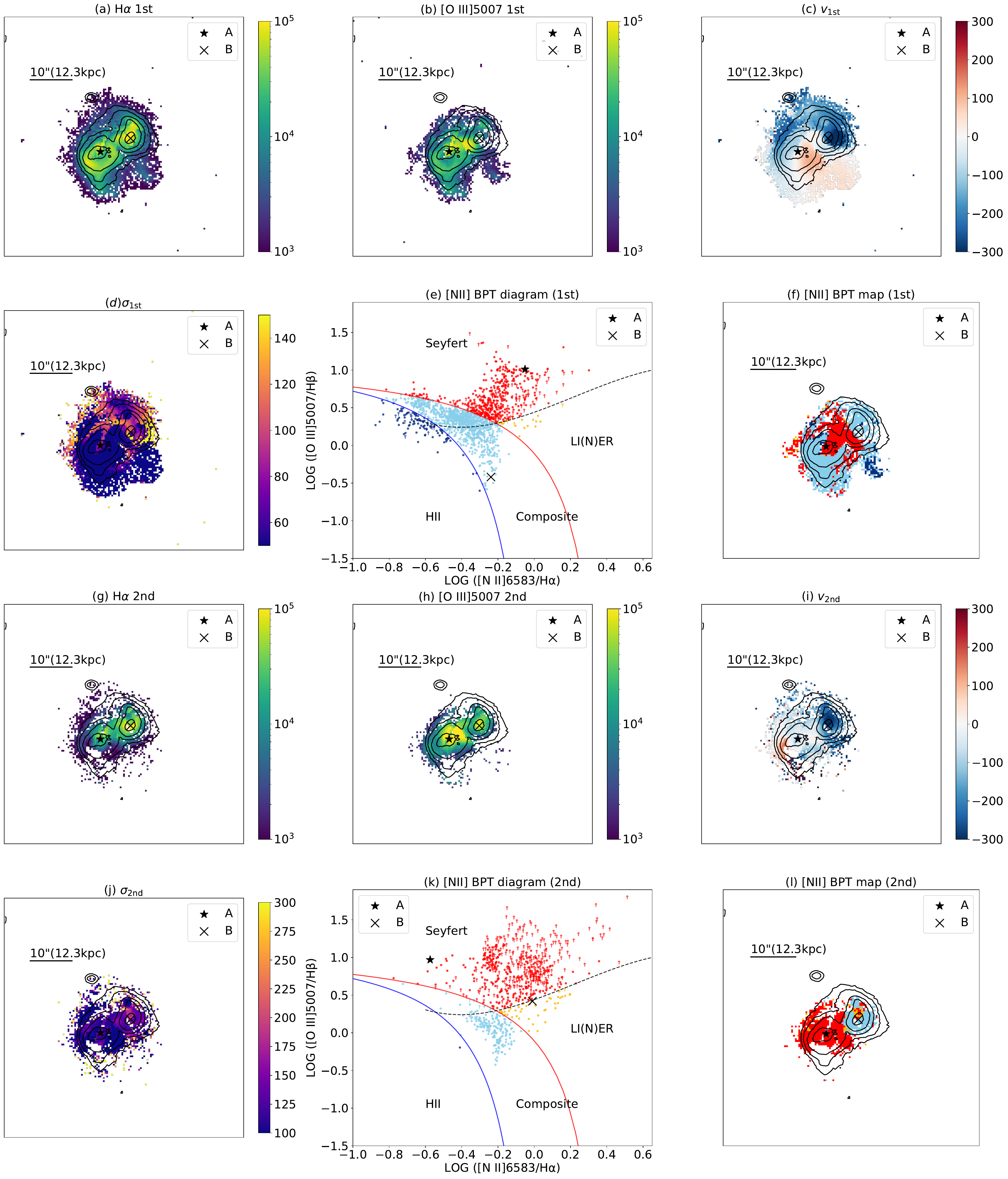}
\caption{Surface brightness and kinematic maps of J0853+1626. 
Similar to Figure~\ref{fig:E509_flux_kinematics_bpt}.
The velocities in $v_{\rm 1st}$ and $v_{\rm 2nd}$ maps are relative to the first component of nucleus A. 
}
\label{fig:J0853_flux_kinematics_bpt}
\end{figure*}

\begin{figure*}[ht!]
\includegraphics[width=1\textwidth,trim=0 0 0 0]{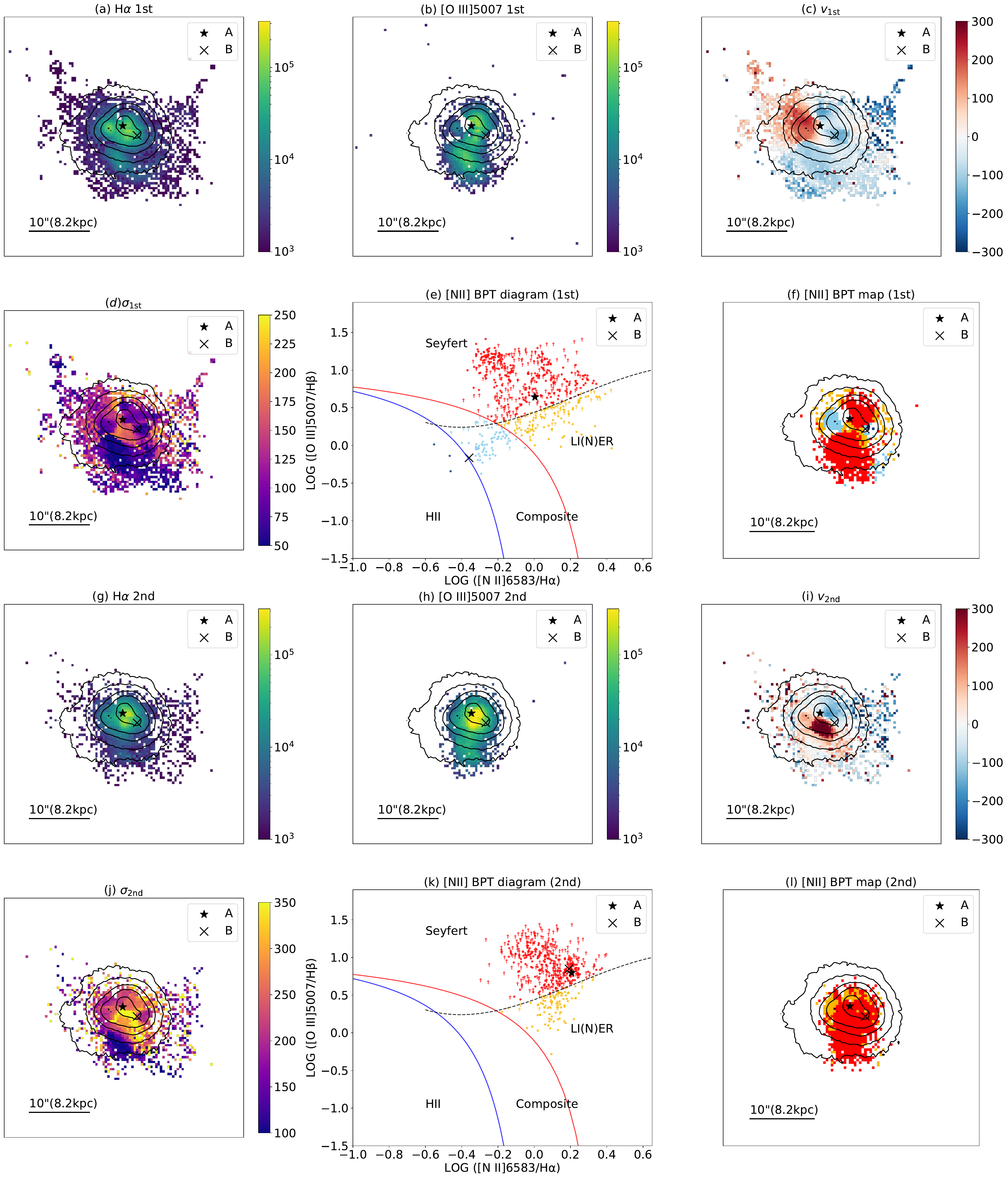}
\caption{Surface brightness and kinematics maps of J0858+1822. 
Similar to Figure~\ref{fig:E509_flux_kinematics_bpt}.
Velocities in $v_{\rm 1st}$ and $v_{\rm 2nd}$ maps are referenced to the first component of nucleus A.
}
\label{fig:J0858_flux_kinematics_bpt}
\end{figure*}

\begin{figure*}[ht!]
\includegraphics[width=1\textwidth,trim=0 0 0 0]{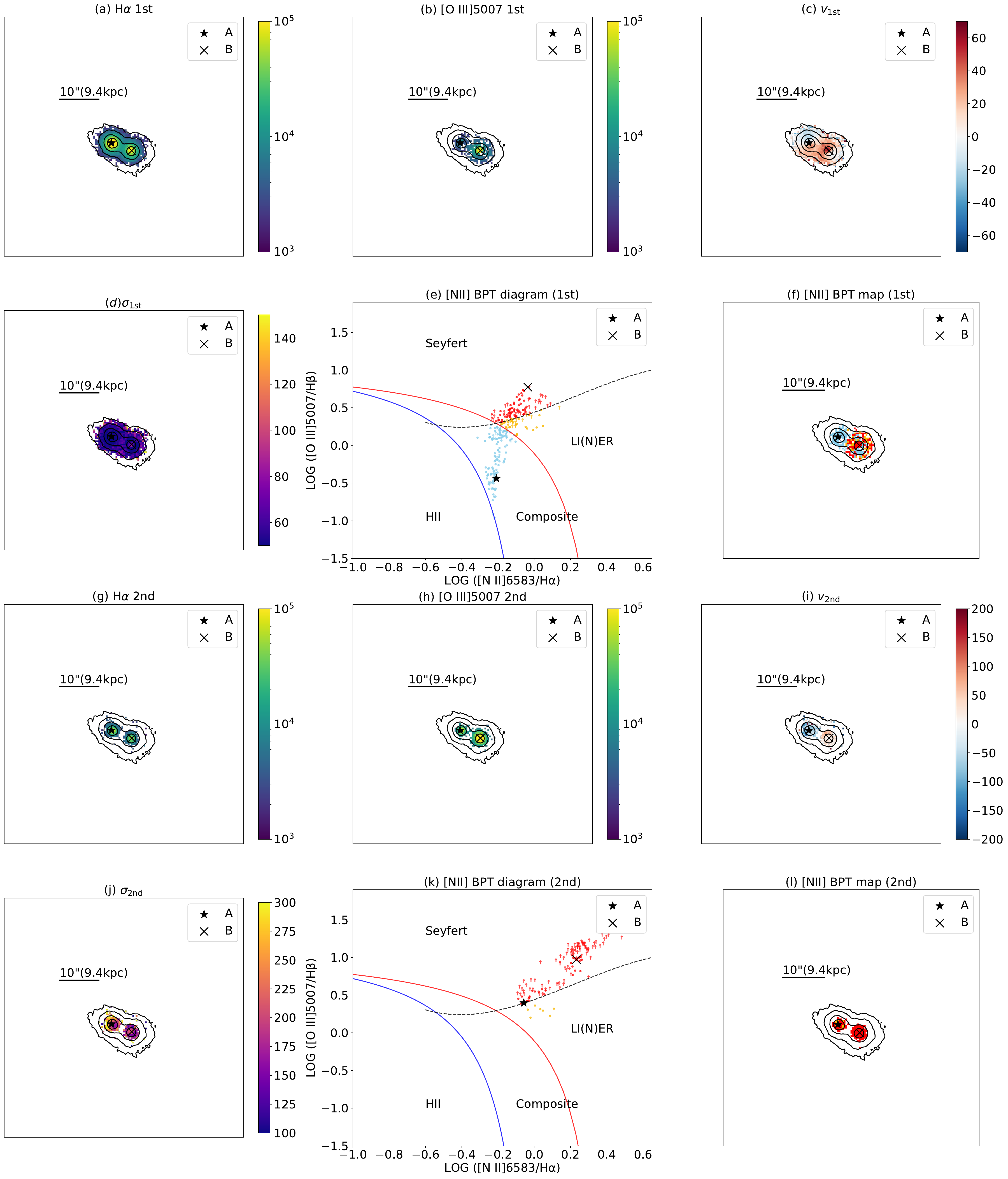}
\caption{Surface brightness and kinematics maps of J1414$-$0000. 
Similar to Figure~\ref{fig:E509_flux_kinematics_bpt}.
Velocities in $v_{\rm 1st}$ and $v_{\rm 2nd}$ maps are referenced to the first component of nucleus A.
}
\label{fig:J1414_flux_kinematics_bpt}
\end{figure*}

\begin{figure*}[ht!]
\includegraphics[width=1\textwidth,trim=0 0 0 0]{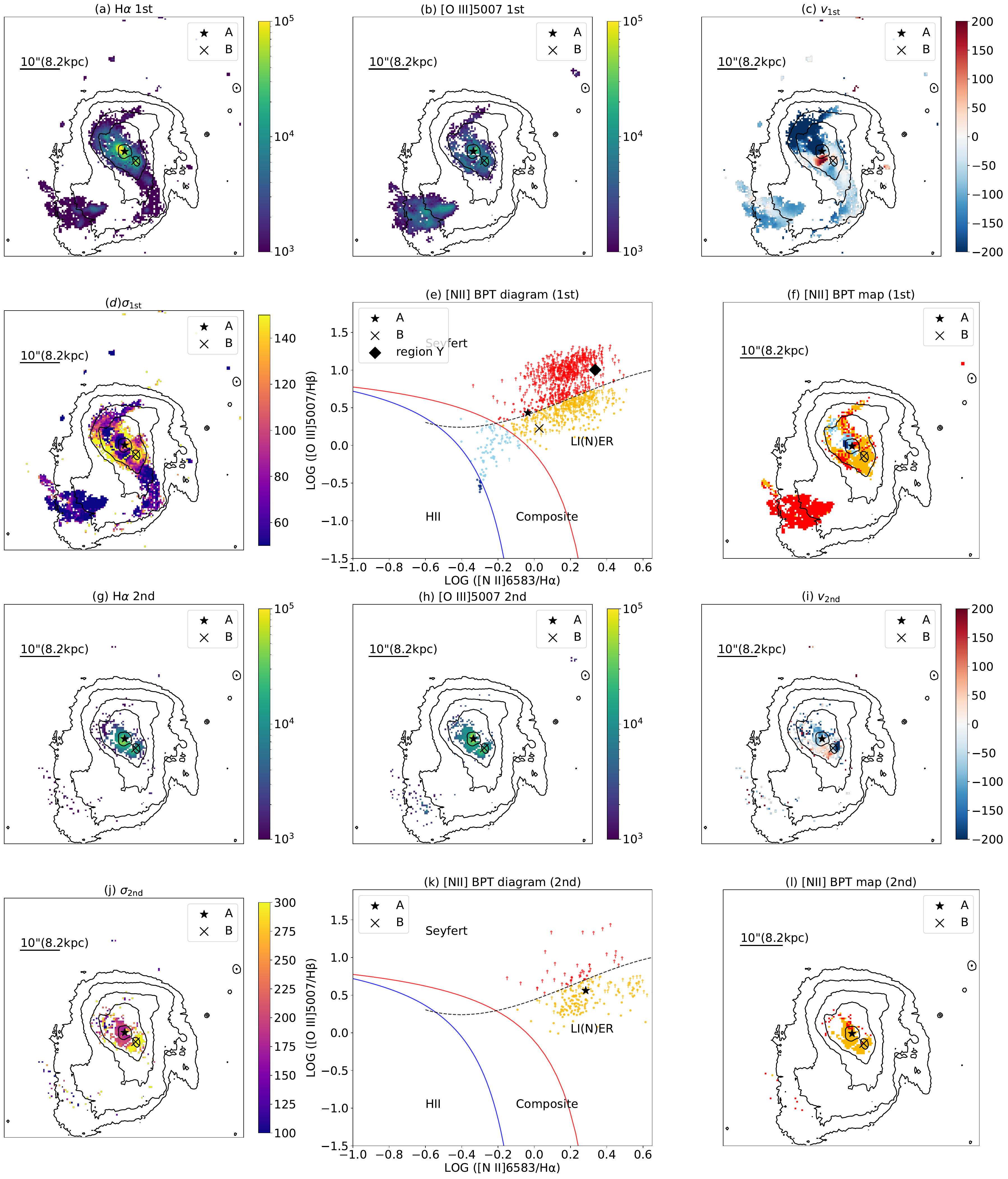}
\caption{Surface brightness and kinematics maps of J1544+0446. 
Similar to Figure~\ref{fig:E509_flux_kinematics_bpt}.
Velocities in $v_{\rm 1st}$ and $v_{\rm 2nd}$ maps are referenced to the first component of nucleus B.
}
\label{fig:J1544_flux_kinematics_bpt}
\end{figure*}

\section{Kinematics and emission line luminosity of nuclei} \label{sec:info_nuclei}

Table~\ref{tab:kinematics-flux table} presents the kinematics, H$\alpha$ and [O~{\sc iii}] luminosities of each nucleus.

\begin{table*}[]
\centering
\begin{tabular}{ccccccc}
\hline
\hline
name        & $v_1 \ (\rm km\,s^{-1})$ & $v_2 \ (\rm km\,s^{-1})$ & $\sigma_1 \ (\rm km\,s^{-1})$ & $\sigma_2 \ (\rm km\,s^{-1})$ & $L_{\rm H\alpha} \ (\rm erg\,s^{-1})$    & $L_{\rm [O\,III]} \ (\rm erg\,s^{-1})$ \\ 
(1) & (2) & (3) & (4) & (5) & (6) & (7) \\
\hline
\hline
ESO 509-066A & -599       & -590       & 76     & 142    & $5.4\times 10^{39}$/$2.1\times 10^{39}$   & $9.3\times 10^{39}$/$6.3\times 10^{39}$        \\ 
ESO 509-066B & -265       & -377       & 161.5  & 534.1  & $7.9\times 10^{40}$/$5.2\times 10^{40}$ & $3.3\times 10^{41}$/$3.8\times 10^{41}$       \\ \hline
IC 1623A     & 21.8       & 31.1       & 74.8   & 178.7  & $1.0\times 10^{40}$/$0.8\times 10^{40}$ & $1.1\times 10^{39}$/$0.9\times 10^{39}$       \\ 
IC 1623B     & 83.4       & 78.6       & 50     & 123.7  & $3.9\times 10^{40}$/$3.5\times 10^{40}$ & $1.4\times 10^{40}$/$2.6\times 10^{40}$      \\ \hline
Mrk 463A     & -35        & -284       & 154    & 546    & $7.9\times 10^{41}$/$4.8\times 10^{41}$ & $1.8\times 10^{42}$/$3.3\times 10^{42}$       \\ 
Mrk 463B     & -56        & -110       & 67     & 233    & $1.3\times 10^{41}$/$0.6\times 10^{41}$ & $5.0\times 10^{40}$/$9.9\times 10^{40}$      \\ \hline
Mrk 739A     & -56        & -183       & 123    & 643    & $5.7\times 10^{40}$/$3.4\times 10^{41}$   & $5.1\times 10^{40}$/$3.6\times 10^{40}$        \\ 
Mrk 739B     & 10         & -18        & 50     & 193    & $4.8\times 10^{40}$/$6.8\times 10^{39}$ & $6.6\times 10^{39}$/$1.3\times 10^{40}$       \\ \hline
NGC 1128A    & 118        & 260        & 202    & 608    & $6.2\times 10^{38}$/$4.6\times 10^{38}$   & $1.7\times 10^{39}$/$2.2\times 10^{39}$        \\ 
NGC 1128B    & 498        & 710        & 376    & 682    & $6\times 10^{38}$/$8.2\times 10^{38}$     & $1.8\times 10^{38}$/$7.3\times 10^{38}$        \\ \hline
NGC 3341A    & 137        & 81         & 50     & 146    & $6.4\times 10^{39}$/$5.1\times 10^{39}$   & $0.4\times 10^{39}$/$1.0\times 10^{39}$        \\ 
NGC 3341B    & -44        & -104       & 58     & 162    & $4.5\times 10^{39}$/$0.8\times 10^{39}$   & $8.5\times 10^{39}$/$2.3\times 10^{39}$        \\ \hline
NGC 7592A    & 23         & -15        & 51     & 104    & $3.4\times 10^{40}$/$1.6\times 10^{40}$   & $6.7\times 10^{39}$/$5.8\times 10^{39}$        \\ 
NGC 7592B    & 35         & 173        & 108    & 352    & $6.4\times 10^{40}$/$1.8\times 10^{40}$   & $1.9\times 10^{40}$/$5.2\times 10^{40}$        \\ \hline
J0051+0020A & 69.7       & 56.8       & 107    & 153    & $8.5\times 10^{40}$/$1.6\times 10^{40}$ & $5.9\times 10^{39}$/$1.8\times 10^{39}$      \\ 
J0051+0020B & -4.1       & 25         & 60     & 154    & $3.3\times 10^{40}$/$1.7\times 10^{40}$ & $1.3\times 10^{40}$/$0.3\times 10^{40}$      \\ \hline
J0853+1626A & 255        & 213.5      & 50     & 113    & $3.2\times 10^{40}$/$2.1\times 10^{40}$ & $4.8\times 10^{40}$/$8.3\times 10^{40}$       \\ 
J0853+1626B & -4.3       & -3.7       & 84.6   & 147    & $9.3\times 10^{40}$/$5.1\times 10^{40}$ & $0.8\times 10^{40}$/$3.7\times 10^{40}$        \\ \hline
J0858+1822A & 83         & 69         & 150    & 256    & $3.4\times 10^{40}$/$4.9\times 10^{40}$ & $3.0\times 10^{40}$/$1.2\times 10^{41}$     \\ 
J0858+1822B & -3         & 117        & 112    & 300    & $2.1\times 10^{40}$/$1.7\times 10^{40}$ & $1.0\times 10^{40}$/$4.9\times 10^{40}$       \\ \hline
J1414$-$0000A & -18.6      & -86.5      & 51     & 223    & $5.4\times 10^{40}$/$1.0\times 10^{40}$ & $0.2\times 10^{40}$/$1.0\times 10^{40}$       \\ 
J1414$-$0000B & 4          & 2          & 50     & 171    & $2.1\times 10^{40}$/$0.8\times 10^{40}$  & $1.3\times 10^{40}$/$3.4\times 10^{40}$       \\ \hline
J1544+0446A & -77        & -40        & 81     & 183    & $1.9\times 10^{40}$/$0.8\times 10^{40}$ & $3.9\times 10^{39}$/$8.9\times 10^{39}$      \\ 
J1544+0446B & 25         & -39        & 149    & 321    & $6.2\times 10^{39}$/$3.1\times 10^{39}$ & $2.9\times 10^{39}$/$4.3\times 10^{39}$      \\ \hline
\end{tabular}
\caption{Columns: (1) Name of the nucleus. 
(2) Velocity of the first gas component ($v_1$) relative to the systemic redshift. 
We derive this velocity from a circular region with a radius of 1.\arcsec2 centered on the nucleus. 
For galaxy pairs in the same MUSE field of view, we correct the spectra using the same redshift. 
(3) Velocity of the second gas component ($v_2$). 
(4) Velocity dispersion of the first gas component ($\sigma_1$) derived from the nuclear 1.\arcsec2 circular region. 
(5) Velocity dispersion of the second gas component ($\sigma_2$). 
(6) Nuclear H$\alpha$ luminosity derived from the 1.2 arcsec circular region. 
The first value represents the first gas component and the second value represents the second gas component. 
(7) Nuclear [O~{\sc iii}] luminosity defined similarly to column (6).
$^{a}$We corrected the datacubes for these galaxies independently because they are not located in the same MUSE FoV.
}
\label{tab:kinematics-flux table}
\end{table*}


\bibliography{sample701}{}
\bibliographystyle{aasjournalv7}


\end{CJK*}
\end{document}